\documentclass[12pt]{aastex}

\graphicspath{{./}}
\makeatletter
\def\input@path{{./}}
\makeatother

\usepackage{epsfig}
\usepackage{amsmath}
\usepackage{amssymb}
\usepackage{graphicx}
\usepackage{appendix}
\usepackage{natbib}
\usepackage{color}
\usepackage{ulem}
\usepackage[breaklinks=true]{hyperref}

\newcommand{\lSect}[1]{\label{sec:#1}}
\newcommand{\lFig}[1]{\label{fig:#1}}
\newcommand{\lEq}[1]{\label{eq:#1}}
\newcommand{\lTab}[1]{\label{tab:#1}}

\newcommand{\vect}[1]{\mathbf{#1}}
\newcommand{\unit}[1]{\vect{e}_#1}
\newcommand{\pran}[0]{\textrm{Pr}}
\newcommand{\abs}[1]{|#1|}
\newcommand{\nut}{\textrm{Nu}_T}
\newcommand{\num}{\textrm{Nu}_{\mu}}
\newcommand{\add}[1]{#1}
\newcommand{\del}[1]{}

\def\gtaprx {\lower .1ex\hbox{\rlap{\raise .6ex\hbox{\hskip .3ex
	{\ifmmode{\scriptscriptstyle >}\else
		{$\scriptscriptstyle >$}\fi}}}
	\kern -.4ex{\ifmmode{\scriptscriptstyle \sim}\else
		{$\scriptscriptstyle\sim$}\fi}}}
\def\ltaprx {\lower .1ex\hbox{\rlap{\raise .6ex\hbox{\hskip .3ex
	{\ifmmode{\scriptscriptstyle <}\else
		{$\scriptscriptstyle <$}\fi}}}
	\kern -.4ex{\ifmmode{\scriptscriptstyle \sim}\else
		{$\scriptscriptstyle\sim$}\fi}}}
\newcommand{\FIGFF}[2]{\ref{fig:#2}{#1}}
\newcommand{\Figff}[1]{\FIGFF{}{#1}}
\newcommand{\FIG}[2]{Fig.~\FIGFF{#1}{#2}}
\newcommand{\Fig}[1]{\FIG{}{#1}}

\newcommand{\Sectff}[1]{\ref{sec:#1}}
\newcommand{\Sect}[1]{\S~\Sectff{#1}}

\newcommand{\App}[1]{Appendix~\Sectff{#1}}
\newcommand{\Eqref}[1]{\ref{eq:#1}}
\newcommand{\Eqff}[1]{(\Eqref{#1})}

\newcommand{\Equation}[1]{Equation~\Eqff{#1}}
\newcommand{\Equations}[1]{Equations~\Eqff{#1}}
\newcommand{\Eq}[1]{eq.~\Eqff{#1}}

\newcommand{\Tab}[1]{Table \ref{tab:#1}}

\newcommand{\wid}{\columnwidth}
\newcommand{\rot}{0}

\begin{document}

\shorttitle{Fingering Convection in Astrophysics}
\title{Chemical Transport and Spontaneous Layer Formation in Fingering Convection in Astrophysics}
\author{Justin M. Brown}
\affil{Department of Astronomy \& Astrophysics\\University of California Santa Cruz\\201 Interdisciplinary Sciences Building\\ Santa Cruz, CA 95064}
\email{jumbrown@ucsc.edu}
\author{Pascale Garaud}
\affil{Applied Mathematics \& Statistics\\Baskin School of Engineering\\1156 High Street\\Mail Stop SOE2\\Santa Cruz, CA 95064}
\author{Stephan Stellmach}
\affil{Institut f\"ur Geophysik\\Westf\"alische Wilhelms-Universit\"at M\"unster\\Germany}

\begin{abstract}
A region of a star that is stable to convection according to the Ledoux criterion may nevertheless undergo additional mixing if the mean molecular weight increases with radius. This process is called fingering (thermohaline) convection and may account for some of the unexplained mixing in stars such as those that have been polluted by planetary infall and those burning ${}^3$He. We propose a new model for mixing by fingering convection in the parameter regime relevant for stellar (and planetary) interiors. Our theory is based on physical principles and supported by three-dimensional direct numerical simulations. We also discuss the possibility of formation of thermocompositional staircases in fingering regions, and their role in enhancing mixing. Finally, we provide a simple algorithm to implement this theory in one-dimensional stellar codes, such as KEPLER and MESA.
\end{abstract}


\maketitle

\section{Introduction}
\lSect{intro}

The phenomenon of fingering convection (otherwise known as thermohaline convection) occurs in regions of stellar and planetary interiors where the mean molecular weight, $\mu$, increases upwards but which are nevertheless stable to the Ledoux criterion. 
If not for the effects of diffusion, such a system would be stable to small perturbations.  
In the presence of diffusion, however, instability can occur.  
Indeed, when a high $\mu$, high entropy parcel is displaced downward, heat rapidly leaks via thermal diffusion.  
Since compositional diffusion is much weaker, the parcel remains denser than its surroundings and proceeds to sink.  
The instability takes the form of finger-like structures and can significantly increase the flux of heat and composition above that of molecular or radiative diffusion.  
Fingering convection appears in several key problems in astrophysics in which an inverse $\mu$-gradient is observed, the most notable of which are planet infall and ${}^3\textrm{He}$ fusion.

Take for instance the problem of planet infall. Stars with detected planets have higher metallicities than those without \citep[e.g.][]{Santos2001,Fischer2005}.  
The relevant question is then whether this observation results from a superior ability of high-metallicity gas to form planets or from pollution by planet infall.  
In the latter scenario, it is thought that stars accrete planets onto their outer convective zones, raising the metallicity at the surface.  
\citet{Vauclair2004} was the first to argue that fingering convection may play an important role in this problem.  
Because the mean molecular weight in the convective zone rises upon absorbing a planet, the radiative region just beneath experiences an inverse $\mu$-gradient and becomes unstable to the fingering instability.  
\add{This drains the excess metallicity from the convective zone into the radiative interior.} However, the time scale for mixing by fingering convection was until recently unknown and is needed to determine how long the post-infall excess metallicity of the convective zone remains observable.  
Thus, in order to determine whether the planet-metallicity connection is an effect of planet formation or infall, we must quantify mixing by fingering convection.

During the process of ${}^3\textrm{He}$ fusion, in which ${}^3\textrm{He}+{}^3\textrm{He}\to{}^4\textrm{He}+2{}^1\textrm{H}$, the total number of particles increases, so the mean molecular weight decreases \citep{Ulrich1972}. 
Because fusion occurs more frequently at higher temperatures and densities, this reaction preferentially decreases the molecular weight deeper in the star, and builds up an inverse $\mu$-gradient.  
This is thought to occur in red giant branch stars.  
Indeed, these stars are observed to undergo a process known as dredge-up, in which particular isotopes (such as $^{3}\textrm{He}$) produced in the deep stellar interior are advected outwards \citep[e.g.][]{Gilroy1989,Charbonnel1994}.  
Simulations by \citet{Charbonnel1994} have produced surface abundances that agree well with observations for the so-called ``first dredge-up.'' However, she noted that some additional mixing in low-mass red giants is needed to account for the observed carbon isotope ratios. \add{\citet{Eggleton2006} initially suggested that a Rayleigh-Taylor instability caused by the inverse $\mu$-gradient could provide this additional mixing. However, }\citet{Charbonnel2007} later proposed that a fingering instability caused by $^{3}\textrm{He}$ burning was more likely but were unable to quantify the mixing due to the lack of constraints on the efficiency of fingering convection.  
Further progress on both subjects---planetary infall and $^{3}\textrm{He}$ fusion---requires the development of improved mixing theories.  

Although the applications in which we are interested are astrophysical, much of the formalism of fingering convection was first developed to address the phenomenon of salt-fingers in the ocean.  
There, temperature and salt play the roles of the entropy and metallicity, and salt diffuses about 100 times more slowly than temperature.  
\citet{Stern1960} first proposed the theory of a ``salt-fountain'' to explain the persistent, small-scale motions observed in warm, salty water lying above cool, fresh water even though density decreases upwards. 
For a recent review of the field, see \citet{Kunze2003}.  
By contrast with astrophysical fingering, it is possible to run laboratory experiments of fingering convection in salt water and measure its turbulent transport efficiency \citep[e.g.][]{Turner1967,Stern1969a,Schmitt1979b}.  
Unfortunately, none of these laboratory experiments are applicable to the parameter regime relevant to astrophysics.  
Because of this, theoretical models of fingering convection in astrophysics have remained, until recently, mostly phenomenological.  

Several prescriptions have been suggested over the past four decades.  
\citet{Ulrich1972} and \citet{Kippenhahn1980} attempted to constrain the dimensions of fingers using stability arguments and determined the resulting thermal and compositional mixing timescales using dimensional analysis. 
Both these prescriptions have a free parameter: in the former, the free parameter is the ``effective inertia of the flow;'' whereas in the latter, it is the aspect ratio of the fingers.  
Later, \citet{Schmitt1983} used linear theory to predict the ratio of the heat to the compositional turbulent fluxes in astrophysical fingering but could not determine the absolute fluxes directly from this method.  

Only in the last few years have numerical simulations of fingering convection approaching the astrophysical parameter regime become more readily available \citep[e.g.][]{Denissenkov2010,Traxler2011a}.
\citet{Denissenkov2010} ran 2D simulations to measure the aspect ratio of fingers and used the previous literature and a dimensional argument to find a prescription for mixing by fingering convection in red giant branch stars.  
He concluded that while this process could provide some of the required mixing discussed by \citet{Charbonnel2007}, it alone was insufficient to account for the observed abundances of low-mass red giant branch stars.  
\citet{Traxler2011a} presented 3D numerical simulations of fingering convection and generated an empirical fit of their results to propose transport laws for fingering convection in astrophysics.  
Using their mixing model, \citet{Garaud2011} studied the evolution of the surface metallicity of solar-type stars after planet infall and found that fingering convection would transport the material out of the convective zone too quickly to explain the planet-metallicity connection.  
This then suggests that planets can form more easily in high-metallicity proto-planetary disks. \add{However, the results of \citet{Traxler2011a} also confirm the findings of \citet{Denissenkov2010} on the red giant branch star abundances.}

Clearly, much work remains to be done.  
The failure of existing fingering convection models to explain red giant branch star abundances suggests that our understanding of the problem remains incomplete.  
Furthermore, as noted by \citet{Vauclair2012}, the mixing prescription of \citet{Traxler2011a} does not fit their data well for systems which are only weakly stratified.  
Based on these considerations, our work here has several goals.  
First, we shall extend the available experimental datasets by running additional simulations closer to the true astrophysical regime.  
Second, we shall develop a theoretical---rather than empirical---prescription for the turbulent fluxes of heat and composition in fingering convection.  
And finally, we shall investigate other mechanisms of transport in fingering regions that may account for the additional mixing needed in the red giant branch case.  

Several such mechanisms have been proposed in the oceanographic literature.  
Not long after the initial theory of salt fingers was proposed, \citet{Stern1969b} realized that fingering convection had the peculiar property of driving coherent, large-scale gravity waves by an instability he termed the ``collective-instability.'' This was confirmed experimentally by \citet{Stern1969a}.  
Another possible large-scale outcome of fingering convection in the ocean is the generation of thermohaline staircases \citep[c.f. observations in the tropical Atlantic by][]{Schmitt2005}.  
These staircases are stacks of distinct, well-mixed, convective layers separated by thin fingering interfaces.  
Currently, the most promising explanation for staircase formation is the $\gamma$-instability, proposed by \citet{Radko2003} and supported by numerical simulations \citep{Stellmach2011}.  
Both large-scale instabilities mix material much more efficiently than ``homogeneous'' fingering convection in the ocean \citep{Schmitt2005}.  
The $\gamma$-instability and collective-instability theories can be applied to the astrophysical case with minor modifications \citep[for details, see][]{Traxler2011a}.  
However, because the viscous and compositional diffusion scales are minuscule in stars, it is currently difficult to perform full 3D numerical simulations in the astrophysical regime that can resolve both the fingers and these large-scale instabilities.  
From a theoretical point of view, \citet{Traxler2011a} predicted that layers could not form by the $\gamma$-instability in astrophysics, but we now investigate this more thoroughly and also pose the question of whether layers may form by the collective-instability instead.  

In \Sect{model}, we describe our model.  
In \Sect{results}, we present the results of existing and new numerical simulations and compare them to the model described in \citet{Traxler2011a}; we find that the latter does not fit our data as the system approaches overturning convection. In \Sect{theory}, we then provide a new model that more precisely fits the simulations and that is constructed from physical principles rather than empirical fits.
We study the conditions for layer formation in \Sect{layers} and conclude in \Sect{discussion} by discussing this new model and its implications in stellar astrophysics.

\section{Model}
\lSect{model}

As discussed in \citet{Traxler2011a}, the characteristic length scale of the fingering instability in astrophysical objects is much smaller than a pressure scale height, which in turn is generally small enough to ensure that the curvature of the star plays little role in the system dynamics. We therefore use a Cartesian grid ($x$, $y$, $z$), with gravity given by $\vect{g}=-g\unit{z}$ to model a small region of the star. In addition, the velocities within these fingers are much slower than the sound speed of the plasma, so we can treat the fluid with the Boussinesq approximation \citep{Spiegel1960}. We assume that there is a background temperature profile, $T_0(z)$, and a mean molecular weight profile, $\mu_0(z)$, which depend only on $z$. If the region considered is small enough, these background profiles can be approximated by linear functions with constant slopes, $T_{0z}$ and $\mu_{0z}$, respectively. We therefore assume that each profile is comprised of a linear background state and a perturbation, e.g. $T_{\textrm{tot}}=zT_{0z}+T$.

The equation of state for the density perturbation, $\rho$, within the Boussinesq approximation, is
\begin{equation}
\lEq{density}\frac{\rho}{\rho_0}=-\alpha{T}+\beta\mu,
\end{equation}
where $\rho_0$ is the mean density of the region. The coefficients $\alpha$ and $\beta$ are those of thermal expansion and compositional contraction and can be obtained by linearizing the equation of state around $\rho_0$. The remaining governing equations for a Boussinesq system are \citep{Spiegel1960}
\begin{align}
\nabla\cdot\vect{u}&=0 , \\
\frac{\partial{\vect{u}}}{\partial{t}}+\vect{u}\cdot\nabla\vect{u}=-\frac{\nabla{p}}{\rho_0}+g(\alpha{T}-\beta\mu)\unit{z}&+\nu\nabla^2\vect{u},  \\
\frac{\partial{T}}{\partial{t}}+\vect{u}\cdot\nabla{T}+w\left(T_{0z}-T^{\textrm{ad}}_{0z}\right)&= \kappa_T\nabla^2T, \textrm{ and} \\
\frac{\partial\mu}{\partial{t}}+\vect{u}\cdot\nabla\mu+w\mu_{0z}&= \kappa_\mu\nabla^2{\mu},
\end{align}
where $\vect{u}=(u,v,w)$ is the fluid velocity, $p$ is the pressure, $\nu$ is the kinematic viscosity, $\kappa_T$ is the thermal diffusivity, $\kappa_\mu$ is the compositional diffusivity, and $T^{\textrm{ad}}_{0z}$ is the background adiabatic temperature gradient, assumed constant within the considered domain. \add{It should be noted here that $\nu$, $\kappa_T$, and $\kappa_\mu$ are assumed to be constant as additional requirements of the Boussinesq approximation. Other interesting dynamics may arise for variable $\nu$, $\kappa_T$, and $\kappa_\mu$, but these are not addressed here.}

In all that follows, we consider stably stratified regions so that $T_{0z}-T^{\textrm{ad}}_{0z}>0$.  
We then non-dimensionalize the governing equations, taking our length scale to be the anticipated finger width, $d=(\kappa_T\nu/g\alpha \abs{T_{0z}-T^{\textrm{ad}}_{0z}})^{1/4}$ \citep{Stern1960,Kato1966}. We scale time, temperature, and composition as $[t]=d^2/\kappa_T$, $[T]=(T_{0z}-T^{\textrm{ad}}_{0z})d$, and $[\mu]=(\alpha/\beta)(T_{0z}-T^{\textrm{ad}}_{0z})d$. With the Prandtl number defined as $\pran\equiv\nu/\kappa_T$ and the diffusivity ratio as $\tau\equiv\kappa_{\mu}/\kappa_T$, the non-dimensionalized governing equations take the following form:
\begin{align}
\lEq{continuity}\nabla\cdot\vect{u}&=0 , \\
\lEq{navier}\frac{1}{\pran}\left(\frac{\partial{\vect{u}}}{\partial{t}}+\vect{u}\cdot\nabla\vect{u}\right)&=-\nabla{p}+(T-\mu)\unit{z}+\nabla^2{u}, \\
\lEq{temperature}\frac{\partial{T}}{\partial{t}}+\vect{u}\cdot\nabla{T}+w&=\nabla^2T, \textrm{ and} \\
\lEq{composition}\frac{\partial{\mu}}{\partial{t}}+\vect{u}\cdot\nabla\mu+\frac{w}{R_0}&=\tau\nabla^2\mu.
\end{align}

\Equation{composition} introduces a new parameter, $R_0$. This so-called ``density ratio'' is the ratio of the stabilizing entropy gradient to the destabilizing compositional gradient \citep{Ulrich1972},
\begin{equation}
R_0=\frac{\nabla-\nabla_{\textrm{ad}}}{\add{\frac{\phi}{\delta}}\nabla_\mu}=\frac{\alpha(T_{0z}-T^{\textrm{ad}}_{0z})}{\beta\mu_{0z}}.
\end{equation}
If $R_0<1$, the destabilizing compositional gradient dominates, and the system is unstable to the Ledoux criterion. One expects the region to be fully mixed by overturning convection. For $R_0\gg1$, the stabilizing entropy gradient dominates, and the system is stable. The only possible transport is via diffusion. For intermediate values, $1<R_0<1/\tau$, \citet{Baines1969} showed that the system is unstable to fingering convection. It is this region of parameter space to which we now focus our attention. As in \citet{Traxler2011a}, we define a reduced density ratio,
\begin{equation}
\lEq{r}
r=\frac{R_0-1}{{\tau}^{-1}-1},
\end{equation}
which remaps the ``fingering regime'' to the interval of $r\in[0,1]$ regardless of the value of $\tau$.

We model fingering convection numerically, and solve Equations \Eqff{continuity} through \Eqff{composition} using a pseudo-spectral double-diffusive convection code as in \citet{Traxler2011a,Traxler2011b} and \citet{Stellmach2011}. To avoid spurious boundary effects, we apply triply-periodic boundary conditions on all perturbations $T$, $\mu$, and $\vect{u}$. The code uses no sub-grid scale model, so the scale on which energy is dissipated must always be fully resolved. In all simulations described below, unless specifically noted, the computational domain is taken to be a cube with side length $100d$. Since the wavelength of the fastest growing mode ranges from $6d$ to $15d$ in the parameter regime considered in this paper, this implies that there are at least 5 to 10 wavelengths of the fastest growing mode per $100d$. We test the required resolution by inspecting the compositional and vorticity fields, which are in general the least resolved ones given the small values of $\pran$ and $\tau$ selected.

\section{Results}
\lSect{results}

\citet{Traxler2011a} explored a significant sample of parameter space, with $\pran$ and $\tau$ ranging from $1/3$ down to $1/30$, as seen in \Tab{results}. However, they only tested a few different density ratios in the cases with $\pran$, $\tau\sim1/30$. Furthermore, owing to computational limitations, they did not explore the low $R_{0}$ limit, the regime where we believe their prescription is most questionable (see \Sect{intro}). Here, we explore more comprehensively in $R_0$ the parameters chosen by \citet{Traxler2011a} and also run simulations down to $\pran$, $\tau\sim1/100$. This enables us to test the validity of their prescription in more detail. It is currently computationally prohibitive to reduce $\pran$, $\tau$ much lower than this, particularly at low $R_0$, where the simulations become increasingly turbulent.\footnote{We note that simulations at much lower values of $\pran$ and $\tau$ have been reported by \citet{Denissenkov2010}. Even though the latter are two-dimensional, it is unlikely that they are fully resolved. Whether under-resolved simulations yield satisfactory estimates for the fluxes is a different question that we shall address in a subsequent publication. The results from the simulations of \citet{Denissenkov2010} \add{($\num=997.6$ for RGB interiors)} are consistent with our theoretical model \add{($\num=1294.7$ for the same parameters, see \Sect{theory})} within thirty percent\del{, which is expected when comparing 2D and 3D results in fingering convection (see Stern et al. 2001)}.} 

As in \citet{Traxler2011a}, we report our results in the form of the Nusselt numbers, which are defined as the ratio of the total vertical flux to the diffused flux and expressed here in terms of the non-dimensional fields as \footnote{The form in terms of dimensional variables $w$ and $T$ is $\nut=1-\langle{wT}\rangle/\kappa_T(T_{0z}-T^{\textrm{ad}}_{0z})$. Note that this definition of the Nusselt number describes the flux of the potential temperature and not that of temperature. \add{The two are only equal when $T^{\rm{ad}}_{0z}=0$.} For a more complete discussion, see \citet{Mirouh2012}.} 
\begin{align}
\lEq{nudefs}\nut &= 1-\langle{wT}\rangle,\\
\num &= 1-\frac{R_0}{\tau}\langle{w\mu}\rangle.
\end{align}
The vertical turbulent fluxes, $\langle{wT}\rangle$ and $\langle{w\mu}\rangle$, are measured by taking the average of the products $wT$ and $w\mu$ over the entire domain.

\add{When we discuss layer formation, we will also use the ratio of the total non-dimensional fluxes,}
\add{\begin{equation}
\lEq{gammadef}
	\gamma=\frac{-1+\langle{wT}\rangle}{-\frac{\tau}{R_0}+\langle{w\mu}\rangle}=\frac{R_0\nut}{\tau\num},
\end{equation}}
\add{and the ratio of the turbulent non-dimensional fluxes,}
\add{\begin{equation}
	\gamma_{\rm turb}=\frac{\langle{wT}\rangle}{\langle{w\mu}\rangle}=\frac{R_0(\nut-1)}{\tau(\num-1)}.
\end{equation}}

\subsection{Typical and Atypical Simulations}
\lSect{simulations}

\begin{figure}
\includegraphics[width=\wid,angle=\rot]{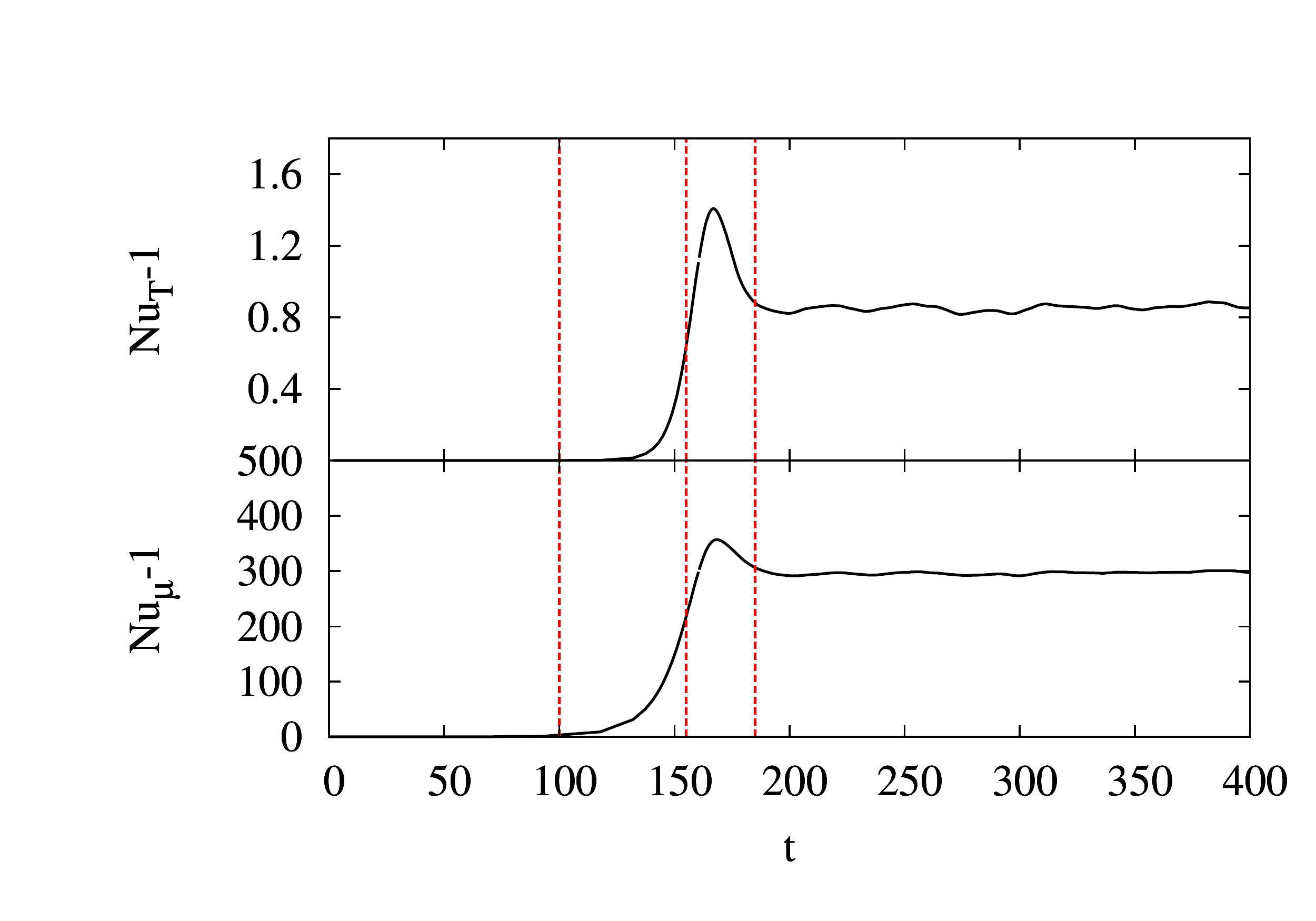}
\caption{\lFig{example} A typical example of the evolution of the thermal (top) and compositional (bottom) Nusselt numbers, $\nut$ and $\num$ (see \Equation{nudefs}) in a simulation with $\pran=1/10$, $\tau=1/30$ with $R_0=3$. Time is measured in units of the thermal diffusion time across a length $d$. We note that the Nusselt curves grow exponentially, peak, and fall to a quasi-steady equilibrium state, which we call the saturated regime. The vertical lines indicate the times of the snapshots in \Fig{fingers}.}
\end{figure}

\begin{figure}
\begin{center}
\includegraphics[width=0.50\wid,angle=\rot]{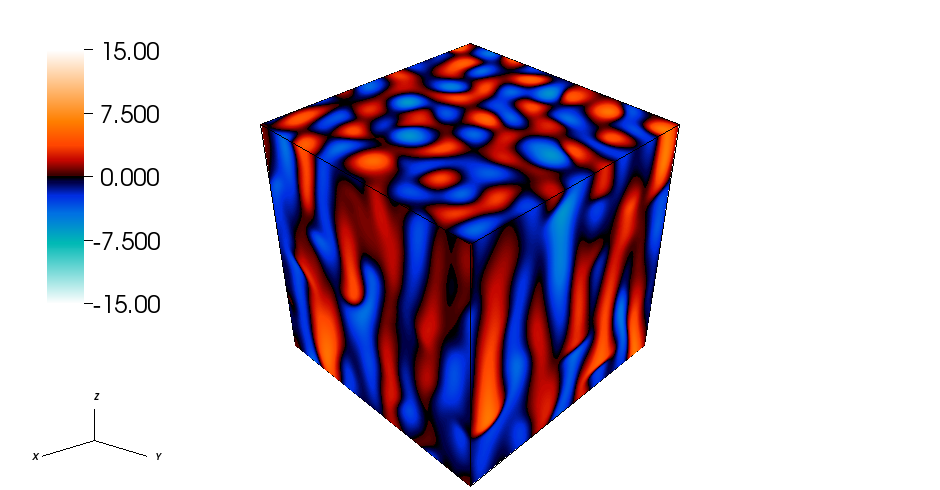}\lFig{elevators} \\
(a) \\
\includegraphics[width=0.50\wid,angle=\rot]{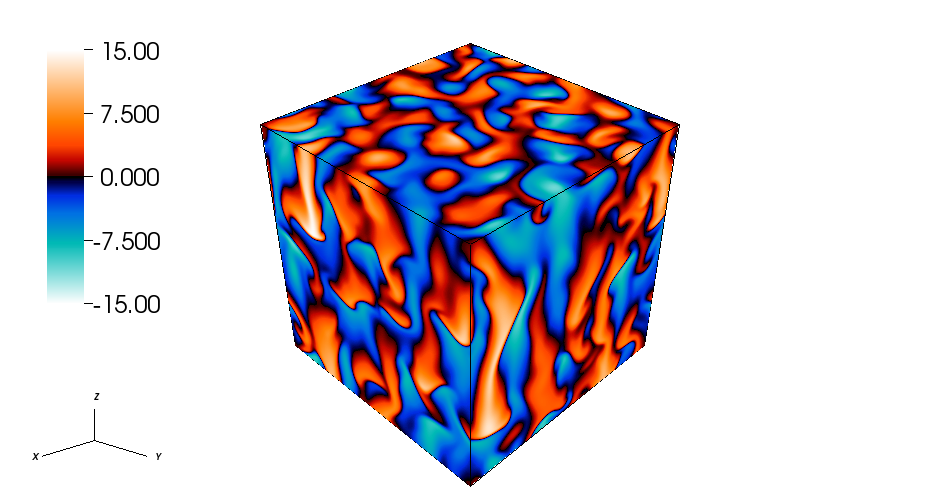}\lFig{peak} \\
(b) \\
\includegraphics[width=0.50\wid,angle=\rot]{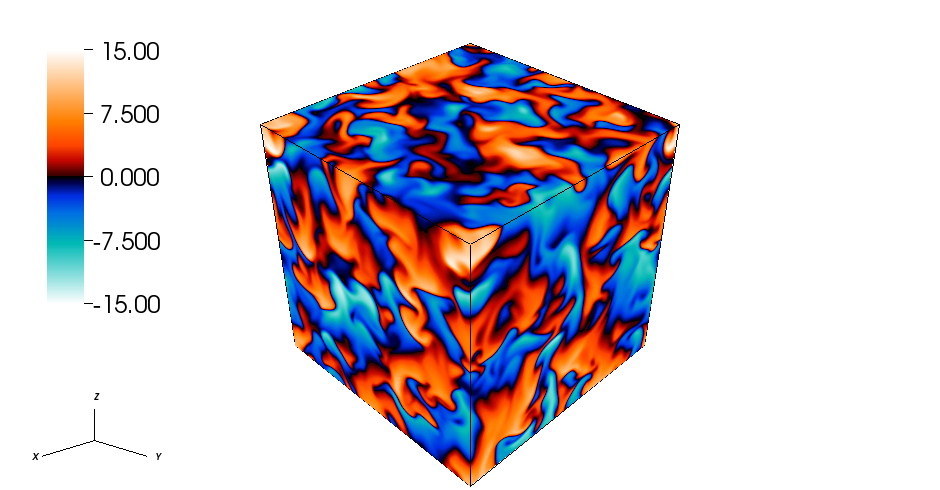}\lFig{saturated} \\
(c) \\
\end{center}
\caption{\lFig{fingers} Snapshots of the compositional perturbation in the simulation shown in \Fig{example} at the characteristic times marked in \Fig{example}. In \Fig{fingers}(a) at $t\approx100$, prior to saturation, the composition field is dominated by tall ``elevator mode'' structures characteristic of linear fingering convection. In \Fig{fingers}(b) at $t\approx155$, though the original elevator modes are still recognizable, they have been disrupted by secondary instabilities. At $t\approx180$, \Fig{fingers}(c) illustrates the ``saturated regime'' discussed in \Fig{example}. Note that the elevator modes have been completely destroyed by the secondary instabilities, leaving a domain filled completely with homogeneous turbulence.}
\end{figure}

In \Fig{example}, we present the temporal evolution of the thermal and compositional Nusselt numbers in a typical numerical simulation of fingering convection, taking $\pran=1/10$, $\tau=1/30$, and $R_0=3$.  
This evolution begins with a period of exponential growth from $t=0$ to $t=160$. The observed growth rate is found to be twice the growth rate of the fastest growing mode according to a linear stability analysis of the governing equations\del{, as expected}. \add{This is as expected, as shown in \Sect{fast}}. The transport peaks around $t=165$, after which the fluxes of temperature and composition saturate (at a level slightly below the peak) and remain roughly constant for the remainder of the simulation.

In order to visualize the saturation process, we present snapshots of the compositional field of the simulation at three stages before and after saturation in Figures \Figff{fingers}(a), (b), and (c), marked with vertical lines in \Fig{example}. In \Fig{fingers}(a), which shows the state just as the primary fingering instability begins to develop, we see tall \del{elevator modes} \add{and thin vertical structures, called ``elevator modes,''} with high $\mu$ plumes sinking and low $\mu$ plumes rising. In \Fig{fingers}(b), which is taken just before the peak in \Fig{example}, we recognize the same vertical structures but notice that smaller-scale instabilities have distorted them. The latter eventually destroy the elevator modes and the system achieves saturation in \Fig{fingers}(c). At this point, very little of the original elevator modes remain discernible, and quasi-steady, homogeneous turbulence dominates the dynamics.

\begin{figure}
\includegraphics[width=\wid,angle=\rot]{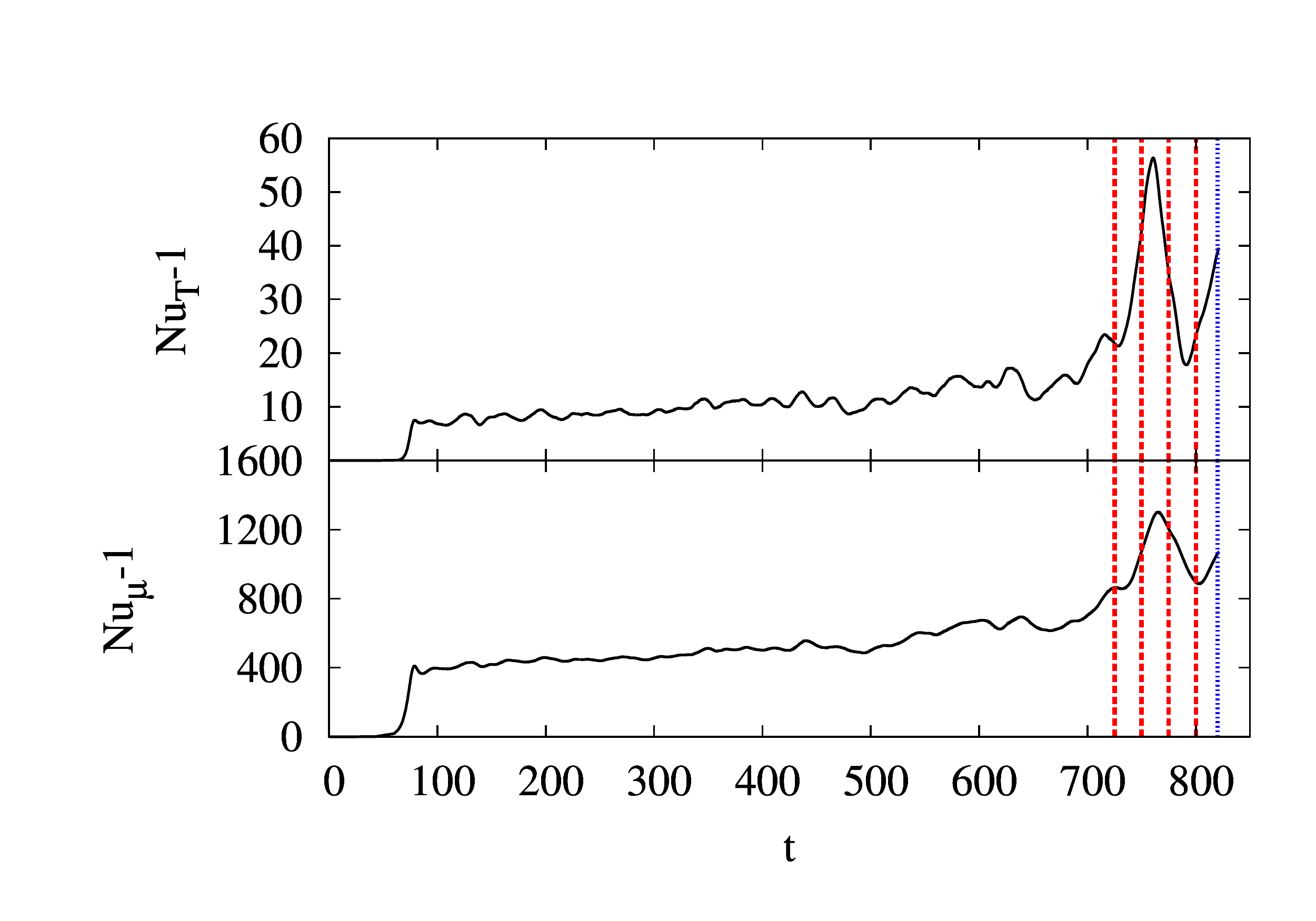}
\caption{\lFig{layers} An ``atypical'' simulation showing the development and effect of large-scale instabilities on the Nusselt numbers, $\nut$ and $\num$ (see \Equation{nudefs}) in a simulation with $\pran=1/10$, $\tau=1/30$, and $R_0=1.1$. We note that this simulation behaves quite differently from the one shown in \Fig{example}. After the initial saturation at $t=80$, both Nusselt numbers gradually increase until $t=700$. They both exhibit quasi-periodic oscillations at twice the buoyancy frequency (though such oscillations are more apparent in the thermal Nusselt number), which suggests the development of gravity waves. At $t=700$, the transport increases dramatically in a manner similar to the onset of layers in the oceanographic case \citep[e.g. simulations by][]{Stellmach2011}. The blue dotted line marks the time at which we plot a snapshot of the compositional perturbation in \Fig{convection}, and the red dashed lines indicate the timesteps for which we calculate the density profiles in \Fig{densitylayer}.}
\end{figure}

While most simulations behave exactly as \Fig{example}, in some cases where $R_0$ is chosen to be very close to the onset of overturning convection (i.e. for small $r$, see \Equation{r}), the Nusselt numbers begin to increase gradually again after saturation. This behavior appears clearly in \Fig{layers}, for which $\pran=1/10$, $\tau=1/30$, and $R_0=1.1$. We distinguish two phases post-saturation: from $t=100$ to $t=600$, the mean Nusselt numbers increase slowly and undergo quasi-periodic oscillations. We find that these oscillations have twice the buoyancy frequency, which is what we expect for gravity waves.~\footnote{\add{Because the velocity, temperature, and composition within low-amplitude gravity waves oscillate at the buoyancy frequency, the turbulent flux (a product of velocity and either temperature or composition) oscillates at twice that frequency.}} Since linear gravity waves do not lead to an increase in the mean flux, the \add{gradual rise of the mean Nusselt numbers} indicates that the waves are highly nonlinear. At $t=600$, the transport suddenly increases dramatically. Meanwhile, the mean density profile in part of the domain inverts, suggesting the formation of a fully convective layer (see \Sect{collective}). This is reminiscent of the formation of gravity waves and layer development in oceanic fingering convection \citep{Stellmach2011}. We will discuss these large-scale instabilities and related transport properties in \Sect{layers}.

\subsection{Extraction of the Fluxes}
\lSect{extraction}

To calculate the saturated fluxes in the homogenous phase of all simulations, i.e. prior to the onset of any large-scale dynamics discussed above, we use the following method.  
We identify the first local minimum in $\nut(t)$ after saturation and set this to be the beginning of the saturated regime. We then fit the remainder of the time series to a linear function of time with a least squares fit, which yields a slope and a $y$-intercept.  
We use regression analysis to estimate an uncertainty for each parameter.  
If the uncertainty in the slope is greater than its magnitude, we then say that the Nusselt curve after saturation is flat and set the end of the saturated regime at the end of the simulation. On the other hand, for a simulation like the one shown in \Fig{layers}, where the flux saturates initially but gravity waves begin enhancing the transport again thereafter, we are only interested in the short interval of time just after saturation. If the linear fit is not sufficiently flat, we reduce the end time of the ``homogeneous phase'' progressively until the fitted slope is zero (within its uncertainty). Once the time interval for homogenous fingering convection has been identified, we calculate the saturated fluxes by averaging them over this interval. The error is taken to be the root mean square of the fluxes.

\begin{deluxetable}{l l l r l r l l}
\tablecaption{\lTab{results} Turbulent flux measurements and respective errors for low $\pran$, low $\tau$ simulations of thermohaline convection}
\tablehead{\colhead{$\pran$} & \colhead{$\tau$} & \colhead{$R_0$} & \colhead{$\nut$} & \colhead{$\sigma_T$} & \colhead{$\num$} & \colhead{$\sigma_{\mu}$} & \colhead{Layers}}
\startdata
1/3 & 1/3 & 1.02003\tablenotemark{a} & $31.0$ & $4.0$ & $120.0$ & $10.0$ & N \\
&& 1.08012\tablenotemark{a} & $23.0$ & $1.0$ & $92.0$ & $5.0$ & N \\
&& 1.10015\tablenotemark{a} & $20.0$ & $1.0$ & $83.0$ & $5.0$ & N \\
&& 1.2003\tablenotemark{a} & $13.5$ & $0.6$ & $60.0$ & $3.0$ & N \\
&& 1.30045\tablenotemark{a} & $9.9$ & $0.5$ & $46.0$ & $2.0$ & N \\
&& 1.50075\tablenotemark{a} & $5.7$ & $0.2$ & $28.3$ & $0.9$ & N \\
&& 2.0015\tablenotemark{a} & $2.16$ & $0.05$ & $9.5$ & $0.3$ & N \\
&& 2.4021\tablenotemark{a} & $1.3$ & $0.01$ & $3.5$ & $0.09$ & N \\
&& 2.8027\tablenotemark{a} & $1.037$ & $0.003$ & $1.33$ & $0.02$ & N \\
&& 2.90285\tablenotemark{a} & $1.0159$ & $0.0009$ & $1.143$ & $0.008$ & N \\
\hline 
1/3 & 1/10 & 1.01 & $28.6$ & $0.4$ & $428.0$ & $6.0$ & Y \\
&& 1.54\tablenotemark{a} & $9.0$ & $0.2$ & $197.0$ & $4.0$ & N \\
&& 3.97\tablenotemark{a} & $1.62$ & $0.02$ & $38.7$ & $1.0$ & N \\
&& 7.03\tablenotemark{a} & $1.075$ & $0.003$ & $7.6$ & $0.2$ & N \\
&& 9.46\tablenotemark{a} & $1.0041$ & $0.0004$ & $1.41$ & $0.04$ & N \\
\hline 
1/10 & 1/3 & 1.10015\tablenotemark{a} & $8.0$ & $0.5$ & $33.0$ & $2.0$ & N \\
&& 1.70105\tablenotemark{a} & $2.16$ & $0.05$ & $8.7$ & $0.3$ & N \\
&& 2.30195\tablenotemark{a} & $1.24$ & $0.01$ & $2.9$ & $0.1$ & N \\
&& 2.8027\tablenotemark{a} & $1.019$ & $0.0007$ & $1.17$ & $0.006$ & N \\
&& 2.90285\tablenotemark{a} & $1.0065$ & $0.0003$ & $1.059$ & $0.003$ & N \\
\hline 
1/10 & 1/10 & 1.003 & $11.36$ & $0.09$ & $169.0$ & $2.0$ & Y \\
&& 1.09\tablenotemark{a} & $8.3$ & $0.7$ & $143.0$ & $9.0$ & N \\
&& 1.45\tablenotemark{a} & $4.3$ & $0.1$ & $86.0$ & $3.0$ & N \\
&& 1.99\tablenotemark{a} & $2.5$ & $0.05$ & $54.0$ & $2.0$ & N \\
&& 2.8\tablenotemark{a} & $1.73$ & $0.02$ & $34.8$ & $1.0$ & N \\
&& 2.98\tablenotemark{a} & $1.62$ & $0.01$ & $31.3$ & $0.6$ & N \\
&& 3.25\tablenotemark{a} & $1.51$ & $0.02$ & $27.4$ & $0.8$ & N \\
&& 4.96\tablenotemark{a} & $1.148$ & $0.003$ & $11.6$ & $0.2$ & N \\
&& 7.03\tablenotemark{a} & $1.032$ & $0.001$ & $3.9$ & $0.1$ & N \\
&& 9.1\tablenotemark{a} & $1.0036$ & $0.0001$ & $1.36$ & $0.01$ & N \\
\hline 
1/10 & 1/30 & 1.1 & $7.6$ & $0.1$ & $500.0$ & $7.0$ & Y \\
&& 1.5 & $4.24$ & $0.06$ & $347.0$ & $5.0$ & N \\
&& 2.0 & $2.77$ & $0.02$ & $252.0$ & $2.0$ & N \\
&& 3.0 & $1.85$ & $0.02$ & $174.0$ & $3.0$ & N \\
&& 6.0 & $1.253$ & $0.004$ & $95.0$ & $1.0$ & N \\
&& 8.0 & $1.1382$ & $0.0006$ & $66.0$ & $0.3$ & N \\
&& 10.963\tablenotemark{a} & $1.059$ & $0.002$ & $32.4$ & $0.7$ & N \\
&& 20.34\tablenotemark{a} & $1.0075$ & $0.0005$ & $7.1$ & $0.4$ & N \\
\hline 
1/10 & 1/100 & 1.1\tablenotemark{b} & 7.4812 & - & 1584.517 & - & N \\
&& 1.3\tablenotemark{b} & 5.1979 & - & 1285.765 & - & N \\
&& 1.5\tablenotemark{b} & 4.1962 & - & 1124.055 & - & N \\
&& 1.7\tablenotemark{b} & 3.581 & - & 1031.832 & - & N \\
&& 1.9\tablenotemark{b} & 2.914 & - & 884.519 & - & N \\
&& 2.1\tablenotemark{b} & 2.6215 & - & 816.522 & - & N \\
&& 2.3\tablenotemark{b} & 2.4437 & - & 789.866 & - & N \\
&& 2.5\tablenotemark{b} & 2.2994 & - & 761.9 & - & N \\
&& 2.7\tablenotemark{b} & 2.0685 & - & 684.936 & - & N \\
&& 2.9\tablenotemark{b} & 1.9298 & - & 636.724 & - & N \\
&& 6.0 & $1.316$ & $0.002$ & $416.0$ & $2.0$ & N \\
&& 12.0 & $1.088$ & $0.002$ & $214.0$ & $4.0$ & N \\
\hline 
1/30 & 1/10 & 1.1 & $4.92$ & $0.08$ & $77.0$ & $1.0$ & N \\
&& 1.5 & $2.26$ & $0.04$ & $37.8$ & $0.9$ & N \\
&& 2.0 & $1.63$ & $0.01$ & $24.8$ & $0.3$ & N \\
&& 3.97\tablenotemark{a} & $1.141$ & $0.005$ & $9.9$ & $0.3$ & N \\
&& 7.03\tablenotemark{a} & $1.021$ & $0.001$ & $2.83$ & $0.09$ & N \\
\hline 
1/100 & 1/100 & 5.0 & $1.072$ & $0.001$ & $102.0$ & $1.0$ & N \\
&& 10.0 & $1.02569$ & $3\times10^{-5}$ & $62.9$ & $0.2$ & N \\
\enddata
\tablenotetext{a}{Data from \citet{Traxler2011a}.}
\tablenotetext{b}{Data from \citet{Radko2012} with a domain size of $38.2\times38.2\times76.3$ and a resolution of $256\times256\times512$. The uncertainties of these fluxes are not know to us.}
\end{deluxetable}

\subsection{Numerical Simulations}

A compilation of our new results with published data from \citet{Traxler2011a} and \citet{Radko2012} is given in \Tab{results}.  
\Fig{compare} shows the corresponding values of the mean Nusselt numbers of both composition and temperature, in the homogenous saturated phase, as a function of the reduced density ratio $r$ (see \Equation{r}). As expected, we see that both $\nut-1$ and $\num-1$ tend to 0 as $r\to1$. In contrast, as $r\to0$, both $\nut-1$ and $\num-1$ increase rapidly, presumably approaching values of Rayleigh-Benard convection in a triply-periodic domain \citep{Calzavarini2005,Garaud2010}. As an additional note, it is clear from \Fig{compare} that $\num$ does not depend on $\pran$ and $\tau$ individually, but rather on their ratio. This effect had already been observed in \citet{Traxler2011a}, and led them to propose their empirical scaling laws:
\begin{align}
  \nut-1&=\pran^{1/2}\tau^{3/2}f(r), \\
  \num-1&=\sqrt{\frac{\pran}{\tau}}g(r),
\end{align}
where $g(r)$ and $f(r)$ have the form $ae^{-br}(1-r)^{c}$, where for $f(r)$, $a = 264\pm1$, $b = 4.7\pm 0.2$, $c = 1.1 \pm 0.1$, and for $g(r)$, $a = 101\pm 1$, $b = 3.6 \pm 0.3$, $c = 1.1\pm0.1$.  

\begin{figure}
\includegraphics[width=\wid,angle=\rot]{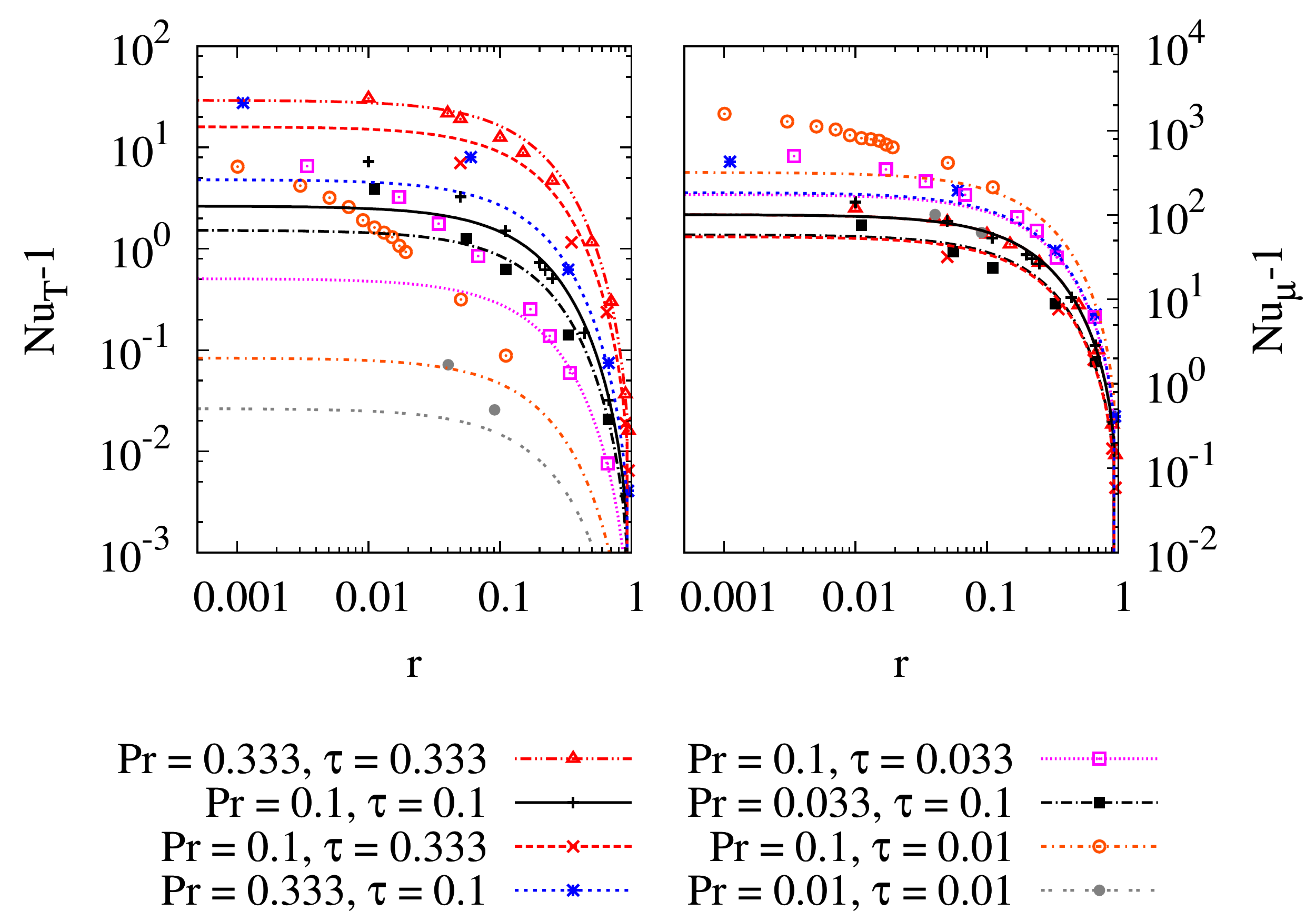}
\caption{\lFig{compare} \add{The Nusselt numbers as a function of the reduced density ratio, $r$, reported from the simulations in \Tab{results} (see references therein). Note that the compositional Nusselt number depends only on the ratio $\tau/\pran$ and $r$, as noted in \citet{Traxler2011a}. For all simulations, the error bars are smaller than the size of the plotted symbols.} \del{This figure contains the same data presented in \Fig{compare} scaled according to the model presented in Traxler et al. (2011a). Note that the scaling works well for large $r$, where the data collapse onto a single curve. However, at low $r$, the scaling breaks down. The points and line styles are the same as those in \Fig{compare}.} \add{We also include the model from \citet{Traxler2011a} as the accompanying curves. It is clear that the model works well for large $r$. However, at low $r$, the model can underestimate the Nusselt number by up to two orders of magnitude.}}
\end{figure}

As discussed in \Sect{intro}, the prescription given by \citet{Traxler2011a} agrees well with the data for high $R_0$; however, these scalings underestimate the transport at low $R_0$, particularly as $\pran$ and $\tau$ decrease, as is illustrated in \Fig{compare}. \del{This is illustrated in \Fig{compare}, in which we present the same data scaled by the suggested prescriptions from Traxler et al. (2011a). For the latter to remain an accurate representation of the data, all the points should lie on a single curve that depends only on $r$.} It is clear from \Fig{compare} that their prescription breaks down at very low $r$, particularly as $\pran$ decreases. For this reason we now propose a theory that better models the data at both large and small $r$. We derive this new theory with physical justification.

\section{Theoretical Model for Fluxes at Saturation}
\lSect{theory}

Recently, \citet{Radko2012} proposed a theory that characterizes the saturation of the fingering instability and predicts the thermal and compositional fluxes at saturation. The fundamental idea of this theory is straightforward: the saturation is thought to be caused by a secondary instability (or ``parasitic'' instability) of the elevator mode and occurs when the growth rate of the secondary instability, $\sigma$, is of the order of the growth rate of the elevator mode, $\lambda$. The latter can be obtained by linearizing the governing equations and deducing the properties of the fastest growing mode. \citet{Radko2012} then calculate the growth rate of the secondary instability using Floquet theory. Unfortunately, their procedure turns out to be computationally quite expensive. In stellar evolution codes, fluxes need to be calculated for different values of $\pran$, $\tau$, and $R_0$ corresponding to each radial mesh point considered at each timestep. Using the \citet{Radko2012} method would be prohibitive for this application.

We pursue a theoretical model that is effectively a simplified form  of the \citet{Radko2012} theory but has the advantage of being much more straightforward and can thus be used in real time in stellar evolution codes. In what follows, we first review the derivation of the growth rate of the fastest growing mode, $\lambda$, in \Sect{fast} \citep{Baines1969}. In \Sect{nusseltmodel}, we then propose a very simple analytical formula for the growth rate of the secondary instability, $\sigma$, and finally determine the Nusselt numbers expected when $\sigma\sim\lambda$.

\subsection{Fastest Growing Modes}
\lSect{fast}

The properties of the fastest growing fingering mode can be determined by linearizing the governing equations and assuming exponential solutions of the form 
\begin{equation}
q=\hat{q}e^{\lambda{t}+i\vect{k}\cdot{\vect{x}}},
\end{equation}
for each of the velocity, temperature, composition and pressure perturbations \citep{Baines1969}. This yields a cubic equation for the growth rate $\lambda$, with coefficients that depend on the wave vector, $\vect{k}$, and the governing non-dimensional parameters, $\pran$, $\tau$, and $R_0$. It can be shown that \add{all properties of} the fastest growing mode \add{(e.g. the velocity, temperature, and salinity fields)} are independent of $z$, hence the term ``elevator'' mode used to describe them. Since our equations are symmetric in $x$ and $y$, $\lambda$ only depends on the magnitude $l$ of the horizontal wavenumber and not on its direction.

The resulting cubic then reads \citep{Baines1969}:
\begin{align}
\lEq{cubic}\lambda^3&+a_2\lambda^2+a_1\lambda+a_0=0\textrm{, where} \\
a_2&=l^2(1+\pran+\tau), \nonumber\\
a_1&=l^4(\tau\pran+\pran+\tau)+\pran\left(1-\frac{1}{R_0}\right), \nonumber\\
a_0&=l^6\tau\pran+l^2\pran\left(\tau-\frac{1}{R_0}\right). \nonumber
\end{align}
To identify the fastest growing mode, we maximize $\lambda$ with respect to the wavenumber $l$. This yields a quadratic for $\lambda$:
\begin{align}
\lEq{quadratic}a_2\lambda^2&+a_1\lambda+a_0=0\textrm{, where} \\
a_2&=1+\pran+\tau, \nonumber\\
a_1&=2l^2(\tau\pran+\tau+\pran), \nonumber\\
a_0&=3l^4\tau\pran+\pran\left(\tau-\frac{1}{R_0}\right). \nonumber
\end{align}

Equations \Eqff{cubic} and \Eqff{quadratic} can be solved numerically simultaneously to determine the wavelength and growth rate of the fastest growing mode. More interestingly for anyone interested in quick analytical estimates, it can be shown that in the limit of $\pran$, $\tau\ll{1}$, 
\begin{equation}
\lambda\approx\left\{
\begin{array}{rl}
\sqrt{\pran} & \textrm{if $r\ll\pran\ll1$} \\
\sqrt{\frac{\pran \tau}{r}} & \textrm{if $\pran\ll{r}\ll1$}
\end{array} \right. ,
\end{equation}
and
\begin{equation}
l^2\approx\frac{1}{\sqrt{1+\tau/\pran}}.
\end{equation}
A detailed derivation of these asymptotic limits (and higher-order terms) is presented in \App{asymptotic}.

\subsection{Estimating the Nusselt Number}
\lSect{nusseltmodel}

\citet{Radko2012} consider all possible sources of secondary instabilities for the elevator modes. Here, for simplicity, we restrict our analysis to instabilities arising from the shear between adjacent elevators and neglect viscosity.  
As we demonstrate below, this yields very satisfactory results, and allows for a much simpler solution to the problem.

The elevator modes are characterized by a vertically invariant flow field of the kind
\begin{equation}
\vect{U}(x,t)= w_0 e^{\lambda t} \sin(lx)\unit{z}  \equiv  w_{\rm E}(t) \sin(lx) \unit{z}\mbox{ , }
\lEq{Ushear}
\end{equation}
where $\lambda$ and $l$ are now strictly defined as the growth rate and horizontal wavenumber of the fastest growing mode, introduced in the previous section. Without loss of generality, we have selected the horizontal phase of the mode to be $\sin(lx)$. This flow field is associated with a temperature and composition field
\begin{equation}
\lEq{Tshear}
T_{\rm E}(x,t)=T_0 e^{\lambda t}  \sin(lx) \mbox{   and   } \mu_{\rm E}(x,t)=\mu_0 e^{\lambda t}  \sin(lx)  \mbox{ , }
\end{equation}
where $w_0$, $T_0$, and $\mu_0$ are related via \Equations{temperature} and \Eqff{composition}:
\begin{eqnarray}
\lEq{lamshear}
\lambda T_0 + w_0 = - l^2 T_0  \Rightarrow  T_0 = - \frac{w_0}{\lambda + l^2} \mbox{ , } \\ 
\lambda \mu_0 + R_0^{-1} w_0 = - \tau l^2 \mu_0  \Rightarrow  \mu_0 = - \frac{R_0^{-1}  w_0}{ \lambda + \tau l^2}   \mbox{ , }
\end{eqnarray}
since elevator modes are exact solutions of the full set of nonlinear equations.  
At early times, the velocity shear, $w_{E}(t)$, is weak and the elevator modes grow unimpeded. Using the definitions of the Nusselt numbers given in \Equation{nudefs} with Equations \Eqff{Ushear}, \Eqff{Tshear}, and \Eqff{lamshear}, we find that 
\begin{eqnarray}
\nut(t) &=& 1 - \frac{1}{L_xL_yL_z} \int_x \int_y \int_z w_{\rm E}(t) T_{\rm E}(t) \sin^2(lx)   dxdydz \nonumber \\
 &=& 1 - \frac{1}{2} w_{\rm E}(t) T_{\rm E}(t) = 1 +  \frac{w^2_0}{2(\lambda + l^2)} e^{2\lambda t} \mbox{ . }
\end{eqnarray}
A similar expression can be obtained for $\num(t)$. \add{This shows that the Nusselt numbers initially grow exponentially with a growth rate that is twice that of the fastest growing mode, as mentioned in \Sect{simulations}, prior to saturation.}

We now consider secondary shearing instabilities on the elevator mode flow. Since $\pran\ll{1}$, we assume that viscosity is negligible. In that case, the growth rate of the shearing modes, $\sigma$, is a simple increasing function of the velocity within the fingers. Since the latter increases exponentially, $\sigma$ also increases, while the growth rate of the fingering instability remains constant. We therefore expect that there will be a time where the two are of the same order of magnitude. At this point, the elevator modes are destroyed and saturation occurs. 

As in \citet{Radko2012}, we neglect the temporal variation of the primary elevator modes when evaluating the growth rate of the secondary shearing modes, and simply write
\begin{equation}
\vect{U}(x,z)= \hat w_{\rm E} \sin(lx)\unit{z} \mbox{ . }
\end{equation}
The growth rate of the shearing instability can be found through dimensional analysis or through Floquet theory \add{(see \App{floquet}). The dimensional argument goes as follows: the only relevant velocity in the system is that of fluid within the fingers, $\hat w_{\rm E}$, and the only relevant length scale is $1/l$, so the only relevant growth rate is $\hat w_{\rm E}l$. Hence,}
\begin{equation}
\lEq{sigma}\sigma = K \hat{w}_{\rm E}l \mbox{ . }
\end{equation}
where $K$ is a universal constant of proportionality.  
By ``universal,'' we imply that this constant is independent of any system parameter ($\tau$, $\pran$, $R_{0}$) or any property of the elevator mode.  
All information about the latter is contained in $l$ and $\hat w_{\rm E}$.  

\del{The dimensional argument goes as follows: the only relevant velocity in the system is that of fluid within the fingers, $\hat w_{\rm E}$, and the only relevant length scale is $1/l$, so the only relevant growth rate is $\hat w_{\rm E}l$. The same expression can be derived more rigorously using Floquet theory, as shown in \App{floquet}.}

Assuming that saturation occurs when the shearing instability growth rate is of the order of the fingering growth rate can be written mathematically as 
\begin{equation}
\sigma  = K \hat{w}_{\rm E}l  = c \lambda,
\end{equation}
where $c$ is independent of the fundamental parameters of the system. This equation uniquely determines the velocity within the fingers at saturation to be
\begin{equation}
 \hat w_{\rm E} = \frac{C\lambda\sqrt{2}}{l},
\lEq{we}
\end{equation}
where $C = c/K\sqrt{2}$ is another universal constant (to be determined).  
\Equation{we} is similar to the one used by \citet{Denissenkov2010} to estimate the effective diffusivity by dimensional analysis.  
Our own model, however, departs from his analysis at this point and produces a result that more accurately fits the results of simulations.  

A given elevator mode with velocity $w_{\rm E} (x,z) = \hat{w}_{\rm E} \sin(lx)$ has a temperature profile $T_{\rm E} (x,z) = \hat{T}_{\rm E} \sin(lx)$ with $\hat{T}_{\rm E} = - \frac{\hat{w}_{\rm E}}{\lambda + l^2}$, for the same reasons that led us to \Equation{lamshear}. Hence, at saturation, 
\begin{equation}
\lEq{nuapprox} {\rm Nu}_T =  1 +  \frac{\hat{w}^2_{\rm E}}{2(\lambda + l^2)} = 1 + C^2 \frac{ \lambda^2 }{l^2 (\lambda + l^2)}
\end{equation}
using \Equation{we}. And by similar arguments, it can be shown that
\begin{equation}
\lEq{nuapproxchem}\num =1  + C^2\frac{\lambda^2}{\tau l^2\left(\lambda+\tau{l^2}\right)},
\end{equation}
where $C$ is the same constant as in \Equation{nuapprox}. \add{The physical interpretation of $C$ is now clearer. Since $C$ is related to the relative importance of the shearing and fingering instabilities at saturation, a large value of $C$ indicates that the shearing instability cannot easily disturb the initial fingers, resulting in a large saturated flux. Conversely, a small value suggests that shear easily destroys fingers, resulting in a small saturated flux.}

We compare this prescription with data from this study and those simulations from \citet{Traxler2011a} and \citet{Radko2012} that approach the astrophysical regime. \add{We fit $C$ by using a chi-squared statistical test and find a best fit at $C=7.0096$.} As illustrated in \Fig{tauless}, for a value of $C=7$ and $\tau\le\pran$, we find that our theoretical results match all simulations remarkably well. It is important to note that, while $C$ needs to be fitted, it is a universal constant and cannot depend on $\pran$, $\tau$, or $R_{0}$. The fact that we are able to use only one value of $C$ to correctly fit the data suggests that the premises of this theory are correct.

In \Fig{taumore}, we show that in the opposite case ($\tau>\pran$) we do not observe the same quality of fit. The full analysis by \citet{Radko2012} is also unable to explain these results, so it is likely that \add{saturation in this case operates somewhat differently}. However, since stars are always in the former situation, the poorness of fit of the latter case is interesting but not relevant to our ultimate purpose.

\subsection{Asymptotic Expansions}
\lSect{asymptotics}

As such, Equations \Eqff{nuapprox} and \Eqff{nuapproxchem} yield the thermal and compositional Nusselt numbers provided $\lambda$ and $l$ are known.  
Calculating these quantities requires the simultaneous solution of a cubic and a quadratic (see \Sect{fast}), which can be done numerically quite easily.  
However, analytical approximations to $\nut$ and $\num$ can also be obtained using asymptotic analysis (see \App{asymptotic}), to find that for $r\ll(\tau,\pran)\ll1$,
\add{\begin{align}
\nut-1&\approx{C}^2\pran\left(1+\frac{\tau}{\pran}\right),\\
\num-1&\approx{C}^{2}\sqrt{\frac{\pran}{\tau}}\sqrt{1+\frac{\tau}{\pran}}.
\end{align}
For the case with $(\tau,\pran)\ll{r}\ll1$ and $\tau/\pran\sim1$,
\begin{align}
\nut-1&\approx{C^2}\pran\tau\frac{1+\frac{\tau}{\pran}}{r}, \\
\num-1&\approx{C^2}\sqrt{\frac{\pran}{\tau}}\sqrt{\frac{1+\frac{\tau}{\pran}}{r}}.
\end{align}
Note that in this case, $\num$ depends only on $\tau/\pran$ and not on $\pran$ or $\tau$ individually, as was noticed by \citet{Traxler2011a}.  
And finally for the case as $r\to1$, $(\tau,\pran)\ll1$,
\begin{align}
\nut-1\approx{C^{2}}\tau^{2}\frac{4}{9}\left(1-r\right)^{2}, \\
\num-1\approx{C^{2}}\frac{4}{9}\left(1-r\right)^{2}.
\end{align}}

\del{Recall that the p for large $r$ in Traxler et al. (2011a) were $\pran^{1/2}\tau^{3/2}$ and $\pran^{1/2}/\tau^{1/2}$, which are close to the two latter cases. Note that the lowest order term for $\nut-1$ scales at best as $\pran$. In astrophysical objects, where $\pran\ll1$,} \add{Let us discuss  these results in more detail. First note that in the case $(\tau,\pran)\ll{r}\ll1$, we find $\nut-1\propto\pran\tau$ and $\num-1\propto\pran^{1/2}\tau^{-1/2}$, and for $r\to1$, we find $\nut-1\propto\tau^2$ while $\num-1$ has no pre-factor dependence on $\pran$ or $\tau$. These scalings are reasonably (though not exactly) consistent with the results of \citet{Traxler2011a}, who found $\nut-1\propto\pran^{1/2}\tau^{3/2}$ and $\num-1\propto\pran^{1/2}\tau^{-1/2}$. The discrepancy is not so surprising given that they did not differentiate between different regimes. Second, note that in astrophysical cases, $\pran\ll1$, which implies that } turbulent heat transport by homogeneous fingering convection is negligible. Such is not the case for $\num-1$, which can become quite large, especially in the limit of $r\to0$. We will discuss the implications of these results more in \Sect{discussion}. Finally, note that these asymptotic formulae should work well for the true astrophysical regime where $\pran$, $\tau\ll1$. However, since none of our simulations are particularly close to this regime, the asymptotic approximation does not compare well with the numerical simulations we have run so far.

\begin{figure}
\includegraphics[width=\wid,angle=\rot]{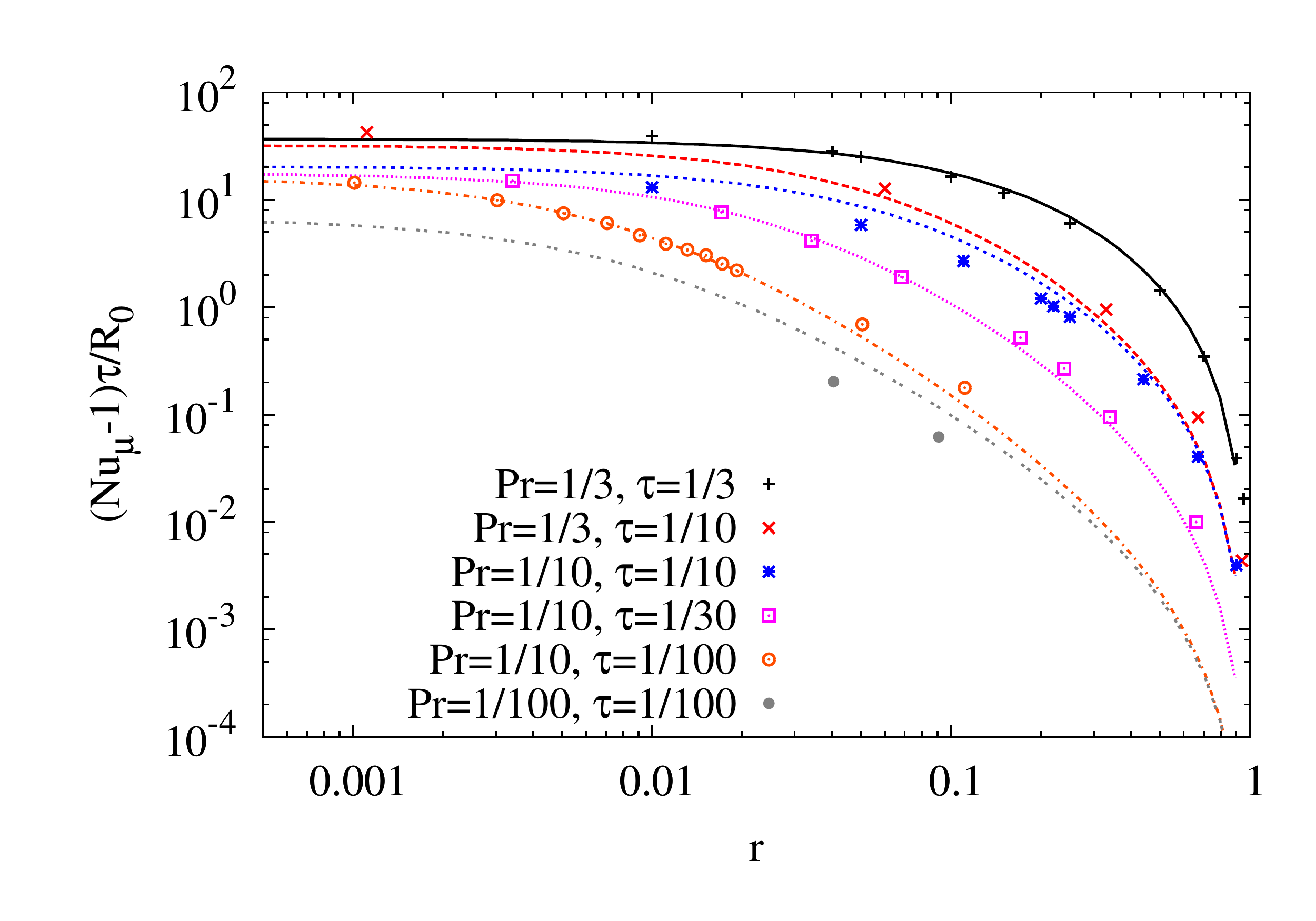}
\caption{\lFig{tauless} Comparison between the turbulent compositional flux observed in simulations (\Tab{results} and references therein), shown as symbols, to our theory (see \Eq{nuapproxchem}), shown as curves, for the simulations with $\tau\le\pran$. Since $\tau$ is always less than or equal to $\pran$ in astrophysics, the cases considered here are physically relevant. The value of the unknown constant $C$ merely shifts the theoretical model predictions vertically by the same amount for all curves, and we find that the best fit has $C\approx7$. Using this value, we find that the measured flux in simulations fits the model remarkably well. We rescale the fluxes by $\tau/R_{0}$ to separate the theoretical curves for ease of viewing.}
\end{figure}

\begin{figure}
\includegraphics[width=\wid,angle=\rot]{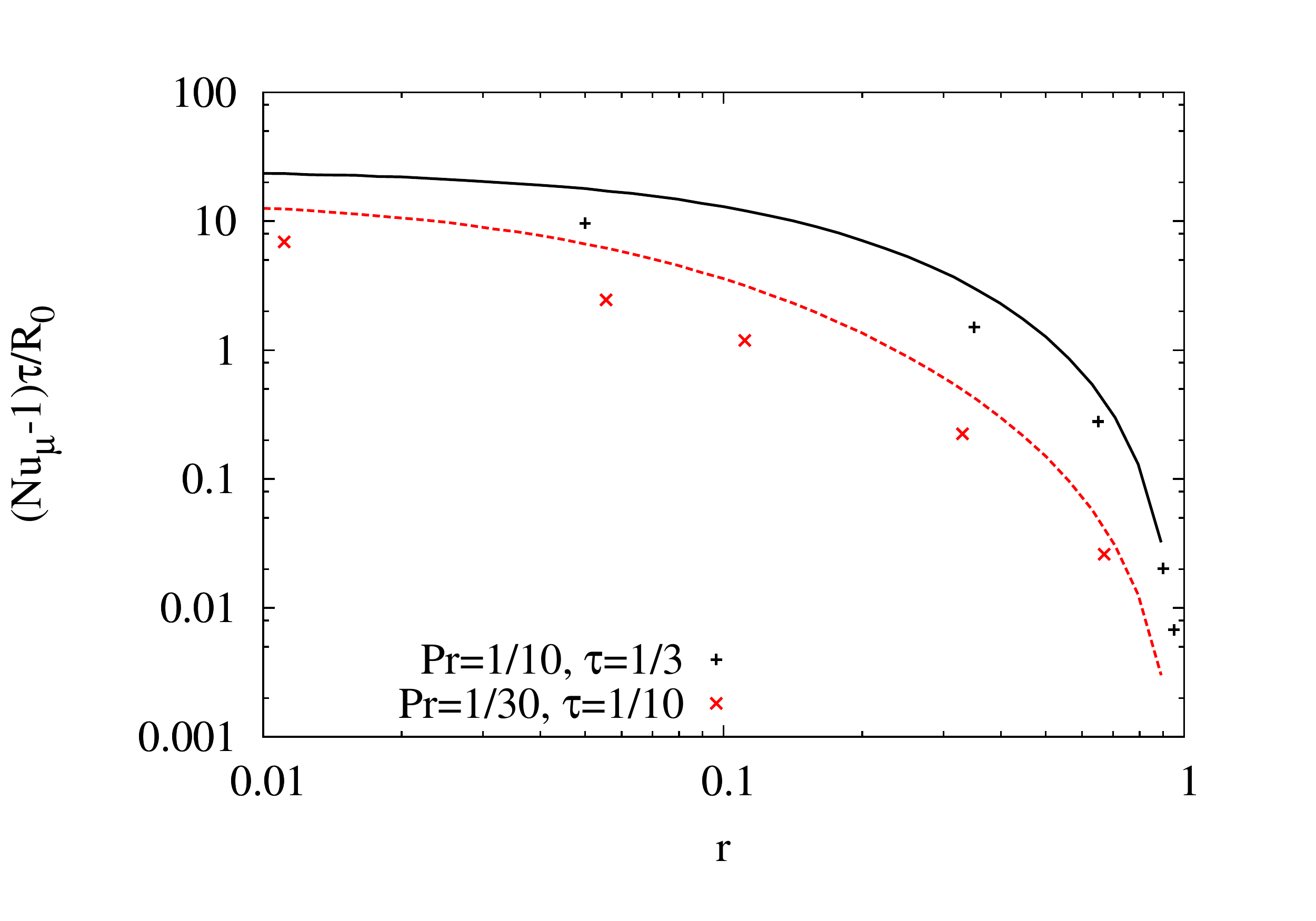}
\caption{\lFig{taumore} Same as \Fig{tauless} but for cases with $\tau>\pran$. The same quality of fit is not observed in these simulations, and the difference between the data and the model is of the order of a few. We rescale the fluxes by $\tau/R_{0}$ to be consistent with \Fig{tauless}.}
\end{figure}

\section{The Development of Layers}
\lSect{layers}

As discussed in \Sect{intro}, substantial evidence from oceanography exists to suggest that mixing rates beyond the levels discussed in the previous section are possible.  
In fingering regions in the ocean, thermohaline staircases---layers of overturning convection separated by thin fingering interfaces---have been observed and have been linked with dramatic increases in the transport of salt and heat \citep[e.g.][]{Schmitt2005}. 
Thus, it is important to consider whether layers can form in the astrophysical case, and whether they also lead to increased transport.  

\subsection{The $\gamma$-instability}

\citet{Radko2003} developed the so-called ``$\gamma$-instability'' theory to explain the formation of thermohaline staircases in fingering regions of the ocean.  
The theory was found to agree with direct numerical simulations of staircase formation in two and three dimensions \citep{Radko2003,Stellmach2011} and has since become well-accepted in the physical oceanography community.  

The original $\gamma$-instability theory is a linear mean-field instability theory.  
\citet{Radko2003} first averaged the governing equations of fingering convection (see Equations \Eqff{continuity} through \Eqff{composition}) spatially, over length scales that span several fingers, and temporally, over multiple finger lifetimes.  
He then performed a linear stability analysis of the averaged, or ``mean-field,'' equations.  
He argued that the stability of the system depends on the quantity $\gamma_{\textrm{turb}}$, defined as the ratio of the turbulent thermal flux, $\nut-1$, to the turbulent compositional flux, $\tau(\num-1)/R_0$.  
More specifically, he showed that homogeneous fingering convection is unstable to layering if and only if $\partial{\gamma_{\textrm{turb}}}/\partial{R_{0}}<0$.  
\citet{Traxler2011a} later improved this theory and showed that in the astrophysical case, this result still holds as long as $\gamma$ is redefined as the ratio of the total fluxes, diffusive and turbulent;\footnote{The discussion of \citet{Denissenkov2010} on layer formation uses $\gamma_{\textrm{turb}}$ rather than $\gamma$, which \citet{Traxler2011a} showed to be incorrect for the astrophysical case.} in other words, a system is unstable to layering if and only if
\begin{equation}
\frac{\partial{\gamma}}{\partial{R_{0}}}<0\textrm{ with }\gamma=\frac{R_{0}\nut}{\tau\num}.
\end{equation}

\begin{figure}
\includegraphics[width=\wid,angle=\rot]{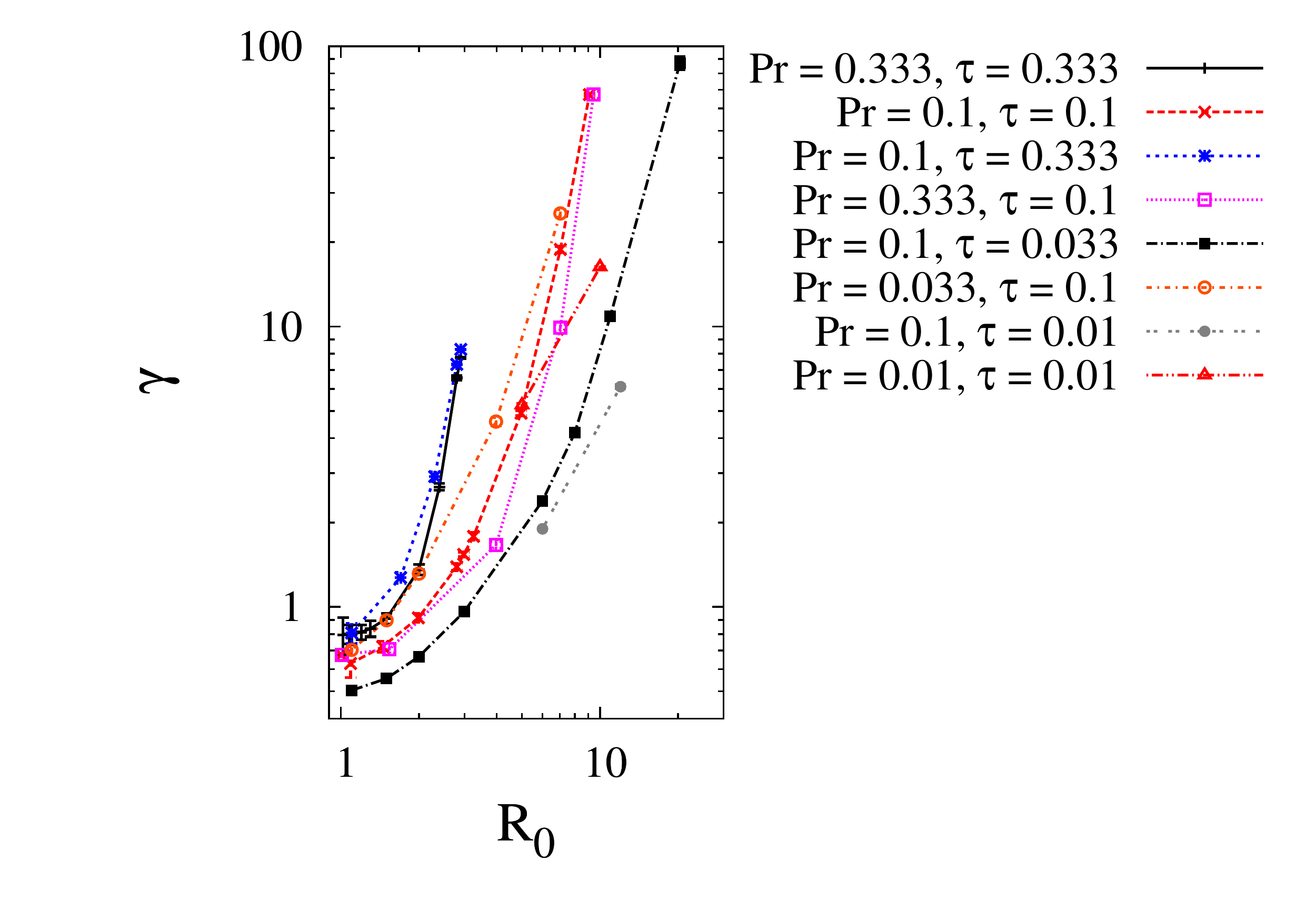}
\caption{\lFig{gammaR0} The total flux ratio, $\gamma$, using the same data as in \Fig{compare}. Note how $\gamma$ increases monotonically with $R_{0}$, which implies that thermocompositional staircases cannot form by the $\gamma$-instability.}
\end{figure}

As mentioned in \Sect{extraction}, some of our simulations at very low $r$ exhibit properties consistent with layering.  
In order to determine whether these layers form via the $\gamma$-instability, we plot in \Fig{gammaR0} the variation of $\gamma$ with $R_{0}$ for various values of $\pran$ and $\tau$.  
Unlike $\gamma_{\textrm{turb}}$, which does decrease for some $r$ at low $\pran$, $\tau$, the quantity $\gamma$ always seems to increase. 
As discussed by \citet{Traxler2011a}, this is because $\nut$ and $\num$ are both dominated by diffusive fluxes at low $\pran$ and $\tau$. \add{The diffusive flux ratio, $R_0/\tau$,} increases strongly with $R_{0}$ \add{because $\tau$ is asymptotically small and $R_0$ approaches $1/\tau$ as the system becomes increasingly stable. This effect dominates the total flux ratio}.  
Since $\gamma$ is always an increasing function of $R_0$, we conclude that the observed layers cannot be forming by the $\gamma$-instability.
Thus, while the $\gamma$-instability can lead to the development of staircases in high $\pran$ oceanic fingering, it is unlikely to be relevant for astrophysical fingering, confirming the results of \citet{Traxler2011a}.  

\begin{figure}
\includegraphics[width=\wid,angle=\rot]{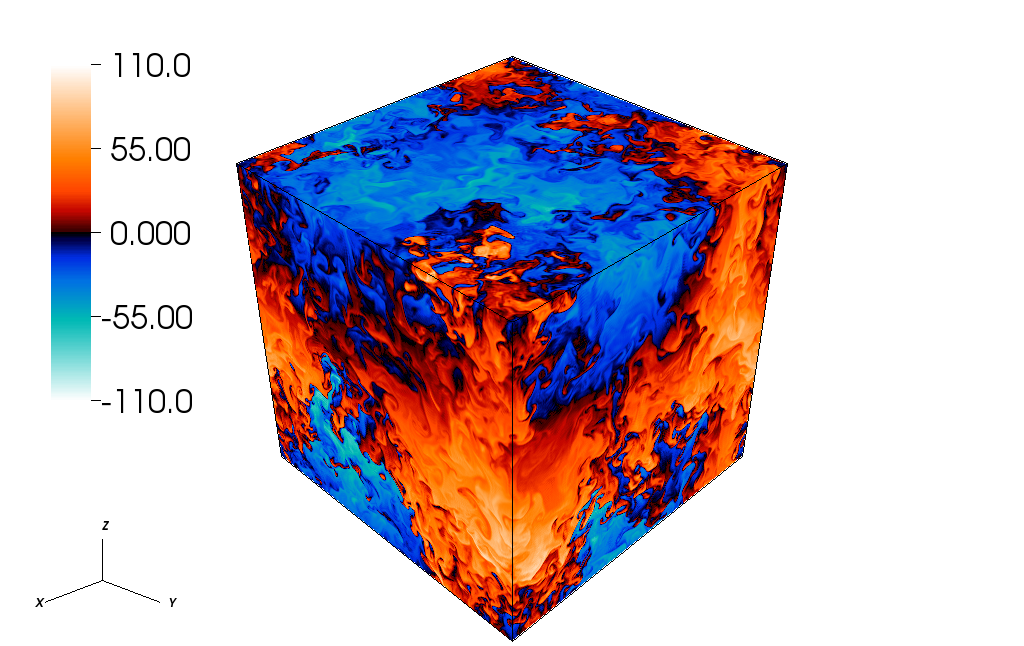}
\caption{\lFig{convection}A snapshot of the compositional perturbation from the same simulation as in \Fig{layers} ($\pran=1/10$, $\tau=1/30$, and $R_0=1.1$) at $t=825$. Large convective plumes are clearly visible and span a significant fraction of the compositional domain. The interface is harder to identify but can be seen near the bottom of the plot, where finger structures are still apparent. Note that it is far from ``flat,'' and instead is highly distorted by the large-scale plumes in the convective layer.}
\end{figure}

\subsection{Simulations with Low $R_0$}
\lSect{collective}

\begin{figure}
\includegraphics[width=\wid,angle=\rot]{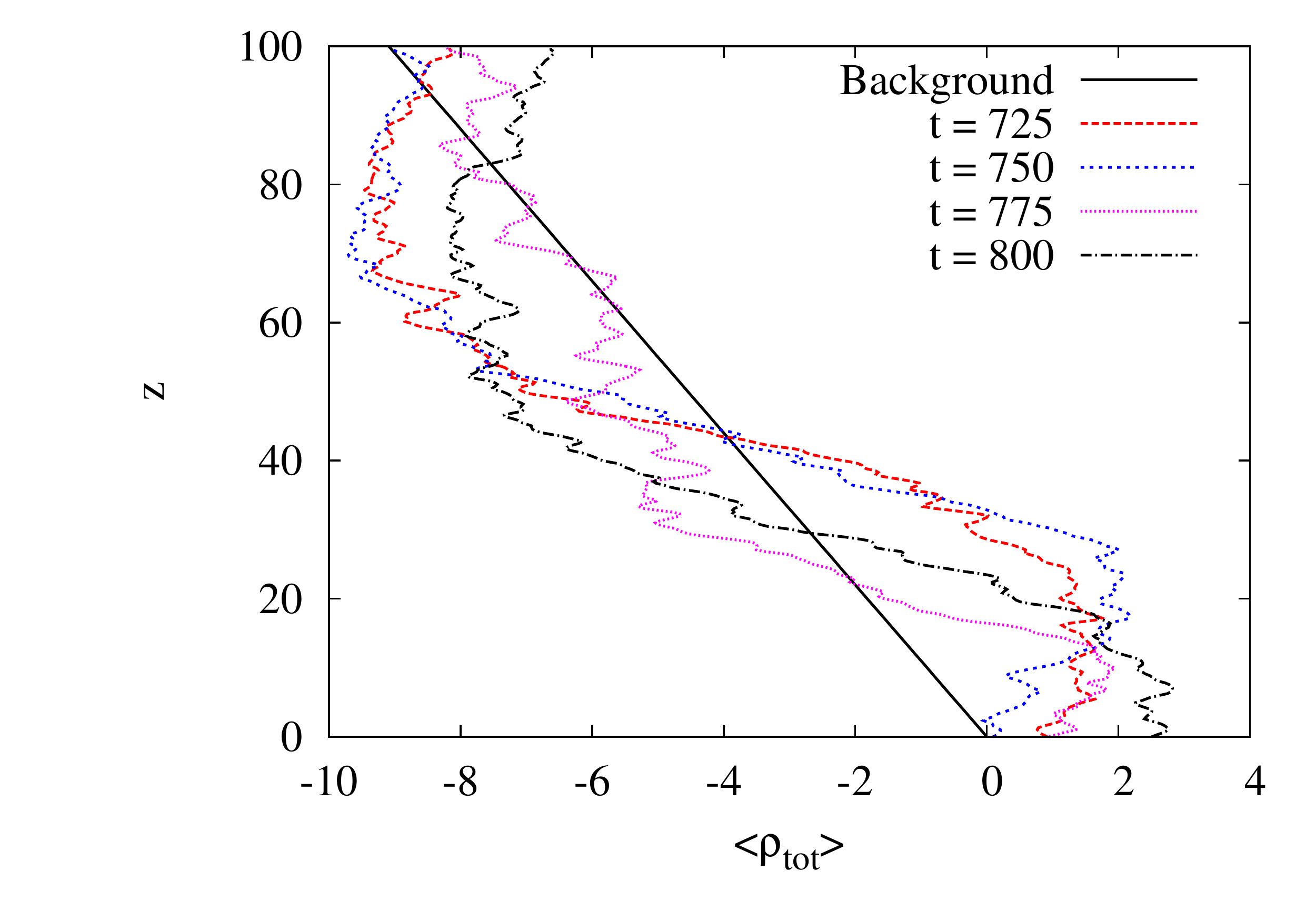}
\caption{\lFig{densitylayer}The horizontally-averaged density profile, $\langle\rho_{\textrm{tot}}\rangle$ (see \Eq{rhotot}) as a function of $z$ for the simulation shown in \Fig{layers} at the four times marked by vertical dashed lines. The background density is shown for reference as a solid line and is always decreasing with height. The sharp density transition in the lower part of the domain is the interface. At times $t=725$, 750, and 800, there exist regions over which the density increases with height, which we call a density inversion. The only time without an inversion, $t=775$, corresponds to the point where the Nusselt number is at its minimum (see \Fig{layers}).}
\end{figure}

Despite the fact that $\gamma$ is a strictly increasing function of $r$, we do observe layer formation in some simulations.  
We now study these results in more detail, focusing on the case presented in \Fig{layers}, which has $\pran=1/10$, $\tau=1/30$, and $R_{0}=1.1$ and shows strong evidence for the formation of a convective layer.  
\Fig{convection} shows a snapshot of the compositional perturbation at $t=825$; large-scale plumes characteristic of overturning convection are clearly visible.  
We can check to see whether these plumes are associated with an inversion of the horizontally-averaged total density profile, which is given by
\begin{equation}
\lEq{rhotot}
\langle\rho_{\textrm{tot}}\rangle=-(1-R_0^{-1})z-\langle{T}\rangle+\langle{\mu}\rangle.
\end{equation}
\Fig{densitylayer} shows $\langle\rho_{\textrm{tot}}\rangle$ as a function of height for the simulation shown in \Fig{layers} at four different times selected during the later stages, where the Nusselt numbers increase significantly.  
The solid line indicates the linear background gradient for reference.  
Note how each of the four density profiles shows a sharp, stably stratified transition region between $z=25$ and $z=45$, the interface.  
Three of the four profiles also have a region where the total density increases with $z$, the convective layer.  
This demonstrates that the homogeneous fingering has indeed transitioned into a layered system with convective layers separated by stable interfaces.  
At $t=725$, shown in \Fig{densitylayer}, the convective layer is just beginning to develop, and a slight inversion of the density profile appears in \Fig{densitylayer}.  
As the simulation approaches the stage of maximum transport at $t=750$ (see \Fig{layers}), the density inverts more strongly.  
The ensuing period of active mixing thickens the interface again and flattens the density profile.  
At $t=775$, the convective mixing briefly shuts down.  
This process repeats, as can be seen at $t=800$, when the density profile once is beginning to invert.  
Note that the interface appears to move up and down during this period, which we attribute to the advection of a large quantity of compositionally dense material by convective plumes.  

This is not the only such simulation in which we observe the formation of staircases: we see this behavior at low $r$ ($\lesssim0.003$) for several values of $\pran$ and $\tau$, including $\pran=1/10$, $\tau=1/10$ and $\pran=1/3$, $\tau=1/10$.  
In all these cases, the $\nut$ and $\num$ time series as well as the properties of the layered state are qualitatively similar to the one shown in \Fig{layers}, \Fig{convection}, and \Fig{densitylayer}.  

Having established that this layering transition cannot be due to the $\gamma$-instability, we look instead at the original mechanism proposed by \citet{Stern1969b} \citep[see also][]{Stern2001} for the formation of thermohaline staircases in the ocean.  
In addition to the $\gamma$-instability, fingering convection is also known to excite large-scale gravity waves through another mean-field instability, the ``collective instability.'' 
\citet{Stern1969b} studied this effect, and showed that large-scale gravity waves can be excited under some circumstances and grow in amplitude exponentially.  
These waves transport momentum, heat, and heavy elements until they begin to ``break,'' at which point the advected material mixes with the background.  
He went on to suggest that if the waves break on large enough scales, this could cause enough local mixing to trigger the formation of staircases.  

In related numerical simulations at oceanographically relevant parameters ($\pran=7$, $\tau=0.3$), \citet{Stellmach2011} found that gravity waves are indeed excited by the collective instability but do not trigger layers.  
Layer formation in their simulations---and thus presumably in the ocean---is caused by the $\gamma$-instability rather than the collective instability.  
Our simulations suggest, however, that the converse may be true in the astrophysical parameter regime.  
As discussed in \Fig{layers}, gravity waves are observed to dominate the dynamics of the system between $t=100$ and $t=600$.  
Furthermore, the gradual increase in $\nut$ and $\num$ during that phase suggests that the waves have high enough amplitudes to interact nonlinearly with each other, and with the background, by contrast with the oceanographic case.  
Indeed, linear waves cause oscillations in $\nut$ and $\num$ without changing their mean values---only nonlinear waves can affect the mean.  
Since the waves in \Fig{layers} are clearly strong enough to interact nonlinearly, we can hope that they may also break and cause mixing on larger scales, following the original hypothesis proposed by \citet{Stern1969b}.  

\begin{figure}
  \centering
  \includegraphics[width=\wid,angle=\rot]{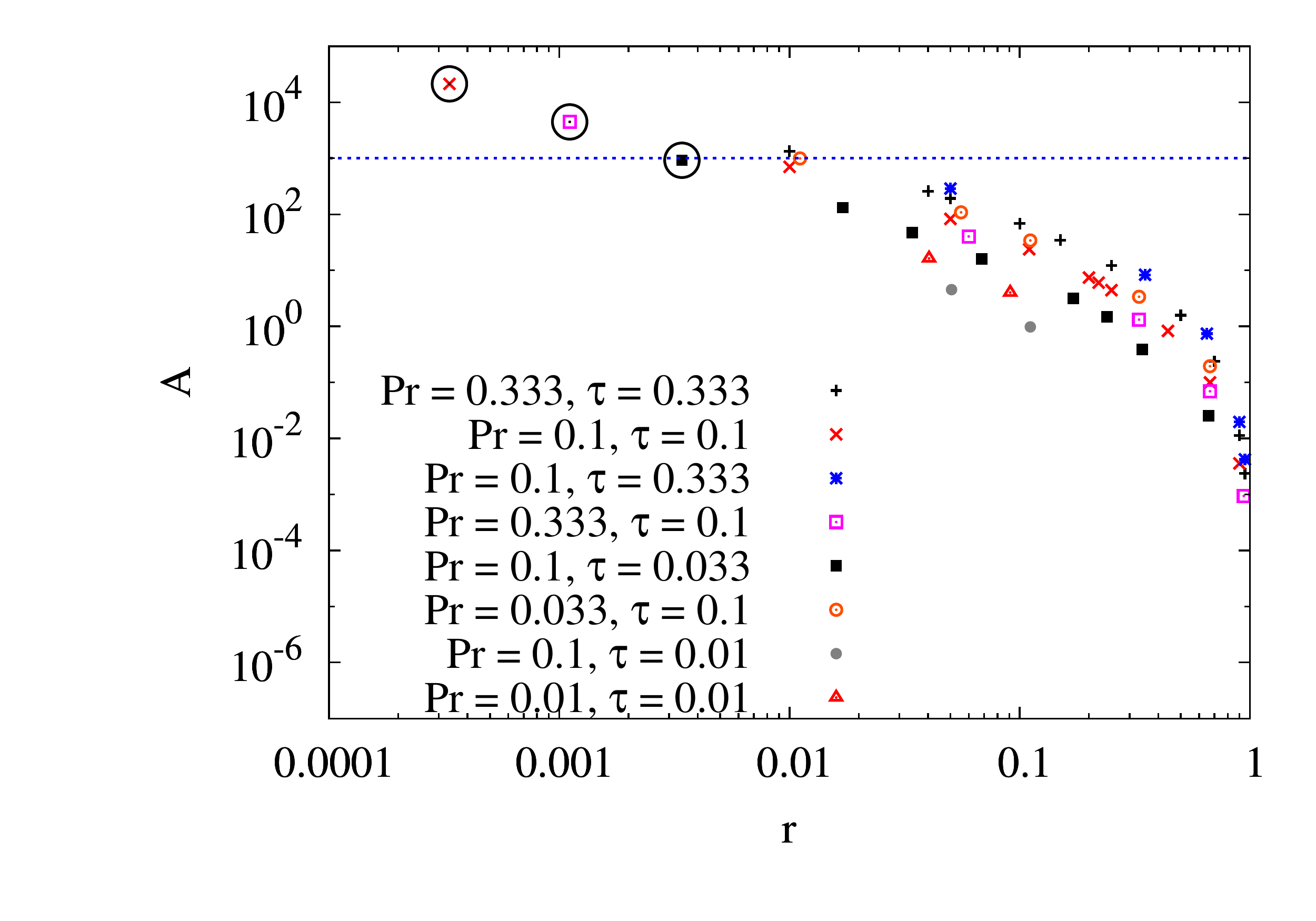}
  \caption{\lFig{resultslayers} Stern number as a function of $r$, calculated from the data presented in \Fig{compare} (and \Tab{results} using equation \Eq{stern}). Simulations that develop into layers are denoted with a large black circle. A Stern number exceeding order unity suggests the possibility of triggering of  gravity wave growth but doesn't necessarily imply layer formation. Here we see that the only cases in which layers form occur when $A\gtrsim10^3$, which is marked with a horizontal dotted line.}
\end{figure}

Checking this hypothesis in detail is difficult with the available simulations.  
However, we can at least determine whether our observed waves do indeed appear in accordance with the collective instability theory.  
A system is unstable to the collective instability if the Stern number, in our notation defined as
\begin{equation}
\lEq{stern}
A=\frac{\left(\nut-1\right)\left(\gamma_{\textrm{turb}}^{-1}-1\right)}{\pran\left(1-R_{0}^{-1}\right)},
\end{equation}
exceeds order unity \citep{Stern2001}.  
Note that this expression uses $\gamma_{\textrm{turb}}=R_{0}(\nut-1)/\tau(\num-1)$ instead of $\gamma$.  
For the simulation shown in \Fig{layers} and \Fig{densitylayer}, $\nut-1=6.55$ and $\gamma_{\textrm{turb}}=0.438$, so that the Stern number is $924$.  
This implies that the system is indeed unstable to the collective instability.  

More generally, we find that all the simulations in which we observe layer formation have Stern numbers of the order of $10^{3}$ or more.  
However, not all such simulations go into layers, as can be seen in \Fig{resultslayers}, so it is likely that the condition for layer formation through the collective instability also depends on the value of $r$.  
It is also possible that these gravity waves grow in all simulations with $A>1$, but since the growth rate depends on the Stern number, gravity waves may be too weak to detect if $A$ is too low.  
If this is the case, it will be difficult to see layer formation in models with smaller $A$ because these simulations are expensive and difficult to integrate for long.  
We discuss this more in \Sect{future}.  

Much work still needs to be done to identify the parameters for which we expect layer formation in the astrophysical regime and to quantify transport in the layered case (see \citet{Wood2013} for related efforts in the case of semi-convection).  
We have only three simulations that develop into layers and thus require a more extensive sample of simulations at low $r$ in order to better understand the conditions of layer formation and their properties as a function of $\pran$, $\tau$, and $r$.  
In every simulation that exhibits layers, we find that a single convective plume spans the majority of the domain and do not see the development of multiple layers, as normally seen in simulations of the fingering regime in the oceanographic case \citep[e.g.][]{Stellmach2011} and in the case of semi-convection \citep{Rosenblum2011,Mirouh2012}.  
To characterize this process, we must increase the size of the domain and extend the integration time to let the layers develop more naturally.  
This will allow us to describe more completely the size scales and transport of thermocompositional layers in astrophysical fingering convection.  
This problem, however, is beyond the scope of this paper and is deferred to a later publication.  

\section{Discussion}
\lSect{discussion}

In the preceding sections, we have used the results from \Sect{results} to formulate a theory that models the Nusselt numbers of homogeneous fingering convection in the astrophysical regime.  
We now provide guidance on how to \del{convert real stellar parameters to the non-dimensional parameters for use in one-dimensional} \add{apply our theory to calculate the effective diffusivities in a} stellar code.  
We quantify the range of parameters where layer formation via the collective-instability may be possible.  
To provide perspective on the impact of these results, we conclude by discussing the predictions from this theory in several astrophysical systems.  

\subsection{A Method for Applying the Model}
\lSect{application}

The first step toward using our model in stellar evolution calculations is to identify the non-dimensional parameters of the system in question at each grid point and/or timestep. Recall that $\pran=\nu/\kappa_{T}$ and $\tau=\kappa_{\mu}/\kappa_{T}$.  
The expressions for the microscopic viscosity and diffusivities can be found in many textbooks of plasma physics \citep[e.g.][]{Chapman1970}.  
In stars, the value of $\pran$ typically ranges from $10^{-7}$ to $10^{-6}$ and that of $\tau$ from $10^{-8}$ down to $10^{-6}$, with $\pran>\tau$.  

The density ratio, $R_0$, is given by
\begin{equation}
R_0=\frac{\alpha{\left(T_{0z}-T_{0z}^{\textrm{ad}}\right)}}{\beta{\mu_{0z}}}=\frac{\nabla-\nabla_{\textrm{ad}}}{\frac{\phi}{\delta}\nabla_\mu},
\end{equation}
where $\nabla=(d\ln{T}/d\ln{P})$ is the actual temperature gradient, $\nabla_{\textrm{ad}}=(d\ln{T}/d\ln{P})_S$ is the adiabatic temperature gradient, $\nabla_\mu=(d\ln{\mu}/d\ln{P})$ is the mean molecular weight gradient, and $\delta=-(\partial\ln{\rho}/\partial\ln{T})_{P,\mu}$ and $\phi=(\partial\ln{\rho}/\partial\ln{\mu})_{P,T}$ \citep[see][for a detailed description of these gradients]{Kippenhahn1990}. The quantities $\nabla$ and $\nabla_{\mu}$ can be calculated from the stellar model grid.  
The other values, $\nabla_{\textrm{ad}}$, $\phi$, and $\delta$, can be calculated from the equation of state.  
Note that $R_{0}$ can take a broad range of values in stellar interiors, but the fluid is fingering-unstable only when $1<R_{0}<\tau^{-1}$.  

\begin{figure}
\includegraphics[width=\wid,angle=\rot]{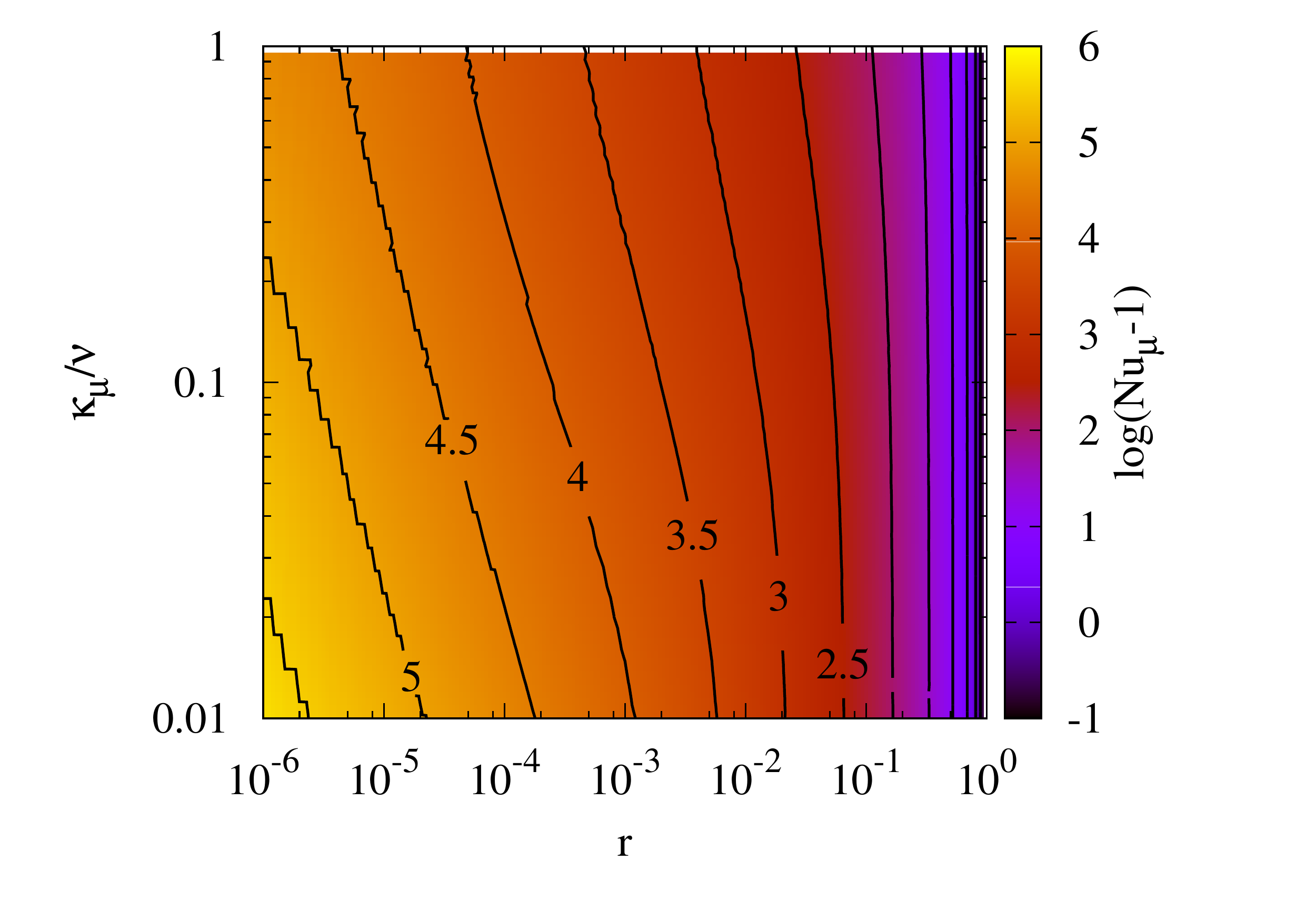}
\caption{\lFig{numcolor} Color plot illustrating the non-dimensional turbulent compositional transport as a function of $\kappa_{\mu}/\nu$ and $r$ from the model presented in \Sect{nusseltmodel}. We assume that $\nu/\kappa_{T}=10^{-6}$ for illustrative purposes. This plot serves as a quick reference for those wanting an order of magnitude estimate of the chemical transport for given $\kappa_{\mu}$, $\nu$, and $r$. Note that the transport becomes dominated by diffusion near $r=1$, as expected. For low $r$, particularly at low $\kappa_{\mu}/\nu$, the transport increases by five to six orders of magnitude.}
\end{figure}

Once $\pran$, $\tau$, and $R_{0}$ are known, one can then calculate $\nut$ and $\num$ according to Equations \Eqff{nuapprox} and \Eqff{nuapproxchem} for a numerically precise answer or by the asymptotic expressions in \Sect{asymptotics} if a rough estimate is all that is needed.  
Since $\num$ is the total flux of composition divided by the background diffusive flux, the effective compositional diffusivity for modeling transport by fingering convection is given by
\begin{equation}
D_{\mu}=\num\kappa_{\mu},
\end{equation}
\add{where}
\add{\begin{equation}
	F_\mu=-D_{\mu}\mu_{0z}
\end{equation}}
\add{is the total dimensional flux of the mean molecular weight through the fingering region.}
An analogous expression can be derived for the heat transport, but this term is nearly always negligible (see \Sect{nusseltmodel}) when $\pran\ll1$ unless layers form.  
We ignore it from here onward.  
By contrast, the turbulent compositional transport can be significant, as can be seen in \Fig{numcolor}, where we plot the logarithm of $\num-1$ as a function of $\tau/\pran=\kappa_{\mu}/\nu$ and $r$, assuming $\pran=10^{-6}$.  
Near the marginal stability limit, $r=1$, the compositional turbulent transport is negligible as well, and $\num-1$ drops to zero.  
However, for low $r$, and particularly for low $\tau/\pran$, the turbulent compositional transport increases by many orders of magnitude.  
This plot is intended as a quick estimate for a large range of parameters; more accurate estimates can be obtained from the expressions given \Sect{nusseltmodel} in \Sect{asymptotics}.  

\begin{figure}
\includegraphics[width=\wid,angle=\rot]{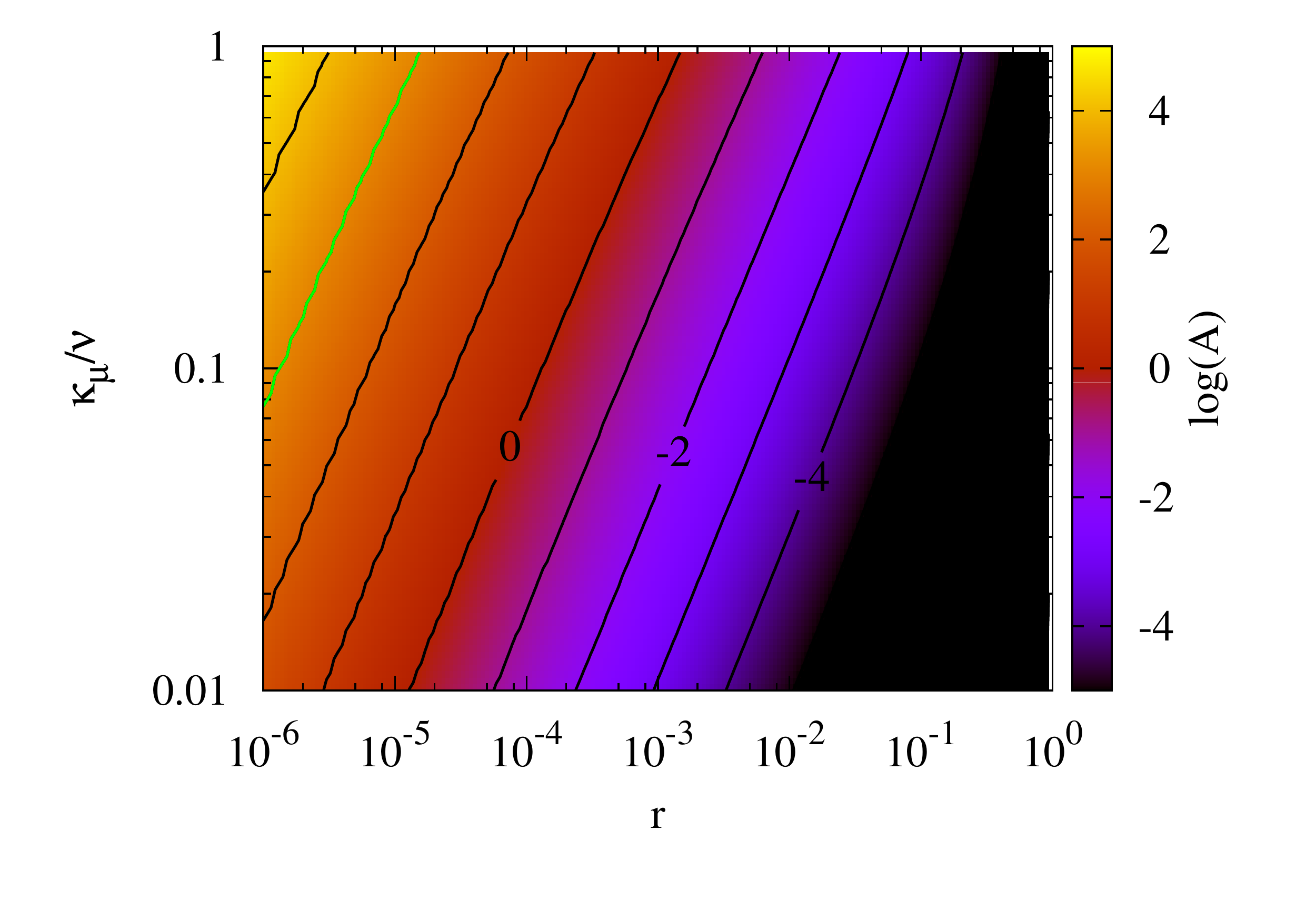}
\caption{\lFig{sterncolor} Color plot illustrating the Stern number as a function of $\kappa_{\mu}/\nu$ and $r$ for the same data as in \Fig{numcolor}. The green contour marks the $10^{3}$ line. If the parameter regime of layers is limited to $A>10^{3}$, the only viable range for layer formation is $r\lesssim10^{-5}$; however, it may be possible for layers to form in other cases, see discussion in \Sect{future}.}
\end{figure}

The results presented so far concern only the case of homogeneous fingering convection.  
As discussed in \Sect{collective}, however, layer formation seems to be possible in some cases and leads to a significant increase in transport compared with the homogeneous levels.  
To determine when this may occur, we use the theory from \Sect{theory} to estimate the Stern number as a function of $\tau/\pran=\kappa_{\mu}/\nu$ and $r$ and show the results in \Fig{sterncolor}.  
We find that the Stern number is largest for small $r$ and large $\tau/\pran$.  
Since we observe layer formation for a Stern number near or above $10^{3}$ (see \Sect{collective}), the same criterion suggests naively that layer formation will only be relevant in stellar interiors (where $\tau/\pran<0.1$) for $r\lesssim10^{-5}$. This is so close to actual overturning convection that it may in fact be irrelevant for practical purposes.  
Of course, as mentioned in \Sect{collective}, much work still needs to be done to confirm these results.  

\subsection{Impact of the New Model on Astrophysical Applications}

We now reexamine several problems that have been recently discussed in the context of fingering convection, namely planet pollution \citep{Vauclair2004,Garaud2011} and the impact of ${}^{3}\textrm{He}$ fusion in red giant branch stars \citep{Charbonnel2007,Denissenkov2010}.  

As first discussed by \citet{Vauclair2004}, fingering convection determines the timescale during which planetary material remains detectable on the surface of a star after infall.  
\citet{Garaud2011} used the fingering transport prescription from \citet{Traxler2011a} to study the problem and determined that this timescale is too short for the excess metallicity of planet-host stars to be related to planet infall.  
Since we have found that the prescription by \citet{Traxler2011a} only underestimates the transport at low $r$ while still recovering the scalings at high $r$, the use of our new, improved model can only increase the turbulent mixing rates and decrease the mixing time, so the qualitative conclusion of \citet{Garaud2011} still holds: the fact that planets are more readily found around high metallicity host stars must be a primordial effect.  

To address the case of red giant branch stars, we compare our results to the numerical simulations of \citet{Denissenkov2010}.  
For his choice of governing parameters, $\pran=4\times10^{-6}$, $\tau=2\times10^{-6}$, and $R_{0}=1700$, he finds $\num=998$.  
He concludes from this result that fingering convection cannot provide all the additional mixing needed in red giant stars.  
Our model finds a remarkably similar value of $\num=1294$ for the same parameters, and thus we confirm his conclusion that additional mixing is necessary to explain observations.  

\subsection{Summary and Future Work}
\lSect{future}

We have developed a simple semi-analytical model for the vertical transport of heat and composition by homogeneous fingering convection that reproduces the results of numerical simulations from this study and from previous studies and can easily be implemented in a stellar evolution code.  
We have found that, under conditions relevant for stellar interiors, the turbulent heat transport is negligible, but compositional transport can be important.  
In particular, this model predicts more efficient transport for weakly stratified systems than was initially suggested in previous work by \citet{Traxler2011a}.  
\del{This model can be extrapolated to stellar parameter regimes and used as a simple method of calculating the compositional turbulent diffusion coefficient in fingering regions of stellar interiors and can be done in real time in a stellar evolution code.} 

We have also found the first evidence for layer formation by fingering convection in the astrophysical parameter regime.  
We have identified the collective instability as the origin of layer formation.  
By contrast, in the oceanographic and semi-convective cases, the $\gamma$-instability causes layer formation \citep[e.g.][]{Stellmach2011,Mirouh2012}.  
Much work still needs to be done to characterize the conditions for layer formation and transport by layered convection in this case.  
With our limited computational resources, we have found that layer formation appears to require high Stern numbers (see \Equation{stern}) to occur.  
If true, this would imply that layers are only likely to form in regions that are very close to being fully convective anyway.  
However, it remains to be seen whether lower Stern numbers can also result in layer formation.  
To study this will require additional simulations and longer integration times than have been covered in this paper. 
In addition to the conditions of layer formation, it is also important to develop a theoretical or empirical model for the transport of this case.  
Our simulations show that transport increases significantly when layers form.  
Concurrent work by \citet{Wood2013} reveals that thermal and compositional transport in the layered semi-convective case scales with the layer height to the power of $4/3$.  
If this scaling also applies here, we may expect that very significant transport rates could occur for large enough layer heights.  
What determines the latter, however, will also require new theories.  

\section*{Acknowledgements}

J.B. and P.G. were funded by NSF grant CBET-0933759 and AST-0807672 and a Regent's Fellowship at UCSC. Part of the computations were performed on the UCSC Pleiades supercomputer, purchased with an NSF-MRI grant. Others used computer resources at the National Energy Research Scientific Computing Center (NERSC), which is supported by the Office of Science of the US Department of Energy under contract DE-AC03-76SF00098. \Fig{fingers} was rendered using ViSiT. P.G. thanks LLBL Hank Childs for his excellent support of the software. \del{This research was supported in part by an award from the Department of Energy (DOE) Office of Science Graduate Fellowship Program (DOE SCGF). The DOE SCGF Program was made possible in part by the American Recovery and Reinvestment Act of 2009.  The DOE SCGF program is administered by the Oak Ridge Institute for Science and Education for the DOE. ORISE is managed by Oak Ridge Associated Universities (ORAU) under DOE contract number DE-AC05-06OR23100.  All opinions expressed in this paper are the author's and do not necessarily reflect the policies and views of DOE, ORAU, or ORISE.} \add{This research is supported in part by the Department of Energy Office of Science Graduate Fellowship Program (DOE SCGF), made possible in part by  the American Recovery and Reinvestment Act of 2009, administered by ORISE-ORAU under contract no. DE-AC05-06OR23100.} We would also like to thank Timour Radko for providing his datasets, which have been included in \Tab{results} and many figures in this paper.

\appendix

\section{Floquet theory for the secondary shearing instability of fingers}
\lSect{floquet}

As discussed in \Sect{theory}, dimensional analysis can be used to show that the shearing growth rate $\sigma$ of the fingering elevator modes must be proportional to the vertical velocity within the finger, $w$, times their horizontal wavenumber, $l$. The same conclusion can be reached using a more formal approach through Floquet theory, detailed in this Appendix.

We loosely follow the method described in \citet{Radko2012}, but use a number of additional simplifications that enable us to obtain analytical solutions. 

We first restrict our analysis to 2D flows. This simplifies the problem greatly, and it can be shown that the final result (i.e. $\sigma \propto w l$) would be the same in 3D. Second, we
only consider the effect of shear in driving secondary instabilities, and neglect that of buoyancy (i.e. we neglect temperature and salinity perturbations). The 2D elevator modes are then well-described 
by the following velocity field: 
\begin{equation}
{\bf U}(x,t)  = w_{\rm E} (t) \sin(lx) \unit{z} \mbox{  .}
\end{equation}
The choice of the phase of ${\bf U}(x,t)$ (i.e. sine or cosine) is arbitrary. We choose the sine for consistency with \citet{Radko2012}. As in \citet{Radko2012}, we then ignore the time-dependence of the elevator mode velocity field ${\bf U}$, setting $w_{\rm E}(t) \equiv \hat{w}_{\rm E}$ where $\hat{w}_{\rm E}$ is constant. Finally, we neglect the effect of viscosity entirely (which is justified in the low Prandtl number astrophysical limit). 

We then consider perturbations ${\bf u}' = (u',0,w')$ to this basic flow, driven by secondary shearing instabilities. As in \citet{Radko2012}, we work with a stream function $\psi$, defined so that $ u' = - \partial \psi/\partial z $ and $ w' = \partial \psi/\partial x $. The linearized momentum equation reads:
\begin{equation}
\frac{\partial {\bf u}'}{\partial t} + {\bf u}' \cdot \nabla  {\bf U} +  {\bf U} \cdot \nabla {\bf u}' = - \pran \nabla p ,
\end{equation}
where the nonlinear term ${\bf U}\cdot\nabla\vect{U}$ naturally vanishes, since the elevator modes are fully nonlinear solutions of the governing equations. Written in terms of the stream function, this becomes:
\begin{equation}
\frac{\partial}{\partial t} (\nabla^2 \psi) + \hat{w}_{\rm E} \sin(lx) \frac{\partial}{\partial z} (\nabla^2 \psi) + \hat{w}_{\rm E} l^2 \sin(lx) \frac{\partial \psi}{\partial z} = 0.
\lEq{psieqshear}
\end{equation}

This is a linear partial differential equation for $\psi$, with coefficients that are constant in time and in the coordinate $z$. We may write the solutions as
\begin{equation}
\psi(x,z,t) = \hat \psi(x) e^{imz + \sigma t} 
\end{equation} 
to satisfy periodic boundary conditions in $z$, as in the original problem. Floquet theory further shows that the $x$-dependence of the solution must be in the form  
\begin{equation}
\hat \psi (x) = e^{iflx} \sum_{n = -N}^N \psi_n e^{inlx},
\end{equation}
where we have used for the sake of clarity the same notation as \citet{Radko2012} for the Floquet term $e^{iflx}$, where $f$ is the Floquet coefficient. Since $f$ needs not be an integer, the 
shearing modes are usually quasi-periodic rather than periodic. Note also that the summation term should in reality be taken from $n=-\infty$ to $+\infty$, but is written here already in its approximate truncated form, in view of a numerical implementation of the problem.  

Plugging this expression into \Equation{psieqshear}, and projecting the result onto individual Fourier modes, we obtain the following tri-diagonal linear system: for each value of $n$ ranging from $-N $ to $N$, we have
\begin{align}
 - \frac{\hat{w}_{\rm E} m}{2} \psi_{n-1} \left[ m^2 - l^2 + l^2 (f+n-1)^2 \right]  - \sigma \psi_n \left[ m^2 + l^2(f+n)^2 \right] + &\nonumber\\
\frac{\hat{w}_{\rm E} m}{2}\psi_{n+1} \left[ m^2 - l^2 + l^2 (f+n+1)^2 \right]  = 0,&
\lEq{diag1}
\end{align}
with the implicit assumption that $\psi_{-N-1} = \psi_{N+1} = 0$. Each secondary mode of instability is identified by the pair $(f,m)$, and grows with rate $\sigma(f,m)$ obtained by setting the determinant of the linear system (\Equation{diag1}) to zero. As in \citet{Radko2012}, we can then scan over all possible values of $m$ and $f$ to find the growth rate of the fastest-growing {\it secondary} instability. 

The beauty of this method is that, by contrast with the more general system studied by \citet{Radko2012}, the scalings for $\sigma$ can be obtained without solving for it at all! Let's define the rescaled vertical wavenumber $\tilde{m} = m/l$ and the rescaled growth rate $\tilde{\sigma} = \sigma / \hat{w}_{\rm E}  l$. The system of equations described by \Equation{diag1} for $n=-N..N$ can then be re-cast in the form   
\begin{align}
 - \frac{1}{2} \psi_{n-1} \left[ \tilde{m}^2 - 1 + (f+n-1)^2 \right]  - \tilde{\sigma} \frac{\psi_n}{\tilde{m}} \left[ \tilde{m}^2 + (f+n)^2 \right] + &\nonumber\\
\frac{1}{2}\psi_{n+1} \left[ \tilde{m}^2 -1 + (f+n+1)^2 \right]  = 0.&
\lEq{diag2}
\end{align}
The rescaled growth rate $\tilde{\sigma}$ is now a function of $\tilde{m}$ and $f$ only, so that maximizing $\tilde{\sigma}$ over all possible values of $\tilde{m}$ and $f$ yields one universal constant. 
In other words, the fastest growing secondary shearing mode satisfies $\tilde{\sigma} = K$, where $K$ is a universal constant, independent of $l$ or $\hat{w}_{\rm E}$. This finally implies, as discussed in the main text, that 
\begin{equation}
\sigma = K \hat{w}_{\rm E} l ,
\end{equation}
where the value of $K$ is somewhat irrelevant since it can be folded into the constant $C$ defined in  \Sect{theory}. 

\section{Asymptotic Analysis}
\lSect{asymptotic}

In Section \Sect{theory}, we derived an analytical expression for the thermal and compositional Nusselt numbers (see \Equations{nuapprox} and \Eqff{nuapproxchem}), in terms of the growth rate $\lambda$ and wavenumber $l$ of the fastest growing elevator mode.  
The latter can be found semi-analytically by solving a cubic and quadratic simultaneously, given by Equations \Eqff{cubic} and \Eqff{quadratic}.  

 While this can be done exactly with a Newton-Raphson relaxation method (or any other numerical method), one may be interested in a quick fully analytical estimate of these fluxes instead. In this Appendix, we derive an approximate formula for the turbulent fluxes, which is increasingly accurate in the asymptotic limit of $\pran$, $\tau \rightarrow$ 0. To do so, we perform an asymptotic analysis of \Equation{cubic} and \Equation{quadratic} in these limits. Since $\tau$ is always of the order the Prandtl number in most astrophysical applications, we simplify our analysis by considering the two limits at the same time, defining the quantity $\phi\equiv\tau/\pran$, and requiring it to be of order unity.

Asymptotic analyses are always much easier to do if one has a good idea of the behavior of the exact solution. The numerical solutions to \Equation{cubic} and \Equation{quadratic} are shown in \Fig{asymplambda} (see data points). We see that the growth rate of the fastest growing mode $\lambda$ as a function of the reduced density ratio $r$ follows different asymptotic laws depending on the range of $r$ considered. For $r \ll (\pran,\tau)$, $\lambda$ appears to be constant and proportional to $\sqrt{\pran}$.  For $r \ll 1$ but not in the previous limit, 
$\lambda$ appears to be proportional to $\pran$, and decreases as $r^{-1/2}$. Finally, in the limit of $r \rightarrow 1$, $\lambda$ drops to 0. We now investigate these three limits in more detail. 

\subsection{First Regime: $r\ll(\pran,\tau)\ll1$}

In the limit of $r\ll(\pran,\tau)$, we find numerically that $\lambda$ is independent of $r$. The same is true for the wavenumber $l$ of the fastest growing mode. This suggests that 
both $\lambda$ and $l$ are continuous in the limit $r \rightarrow 0$, and that one may study this first regime simply by solving \Equation{cubic} and \Equation{quadratic} with $r \equiv 0$ (alternatively, $R_0 = 1$). The resulting equations are (using the definition of $\phi$ given above) : 
\begin{equation}
\lambda^3 + \lambda^2 l^2 (1  + \pran(1+\phi)) + \lambda l^4 \pran (1 + \phi( \pran +1) ) + l^6 \pran^2 \phi + l^2 \pran ( \pran \phi - 1) =0 \mbox{  ,} 
\end{equation}
and 
\begin{equation}
( 1 + \pran(1+\phi)) \lambda^2 + 2 \lambda l^2 \pran (1 + \phi (\pran +1)) + 3 l^4 \pran^2 \phi + \pran   ( \pran \phi - 1)  =0 \mbox{  .}
\lEq{simpquad1}
\end{equation}
Subtracting $l^2$ times the second equation from the first yields the simplified cubic: 
\begin{equation}
\lambda^3 - \lambda l^4 \pran (1 + \phi( \pran +1) ) - 2 l^6 \pran^2 \phi =0 \mbox{  .}
\lEq{simpcub1}
\end{equation}

Next, we study the behavior of $\lambda$ and $l$ with varying $\pran$ and $\tau$ (via $\phi$). We observe that the exact solution for $\lambda$ varies as $\pran^{1/2}$, while $l^2$ is more-or-less independent of $\pran$. Using this information, we propose the following asymptotic expansions: 
\begin{equation}
\lambda = \pran^{1/2}\left( \lambda_0  + \lambda_1 \pran^{1/2}\right)  + \cdots   \mbox{ and } l^2 = l_0^2 + l^2_1  \pran^{1/2}  + \cdots 
\end{equation}
expanding $l^2$ rather than $l$ since \Equation{cubic} and \Equation{quadratic} only depend on the former. 

Plugging these two ansatz into \Equations{simpquad1} and \Eqff{simpcub1}, and solving for all unknowns at their respective orders in $\pran$ yields
\add{\begin{align}
\lEq{asymp1}
& \lambda = \sqrt{\pran}-\pran\sqrt{1+\phi} +  \cdots\mbox{, and} \nonumber\\
& l^2 = \frac{1}{\sqrt{1+\phi}}-\sqrt{\pran}\left(1+\frac{\phi}{(1+\phi)^2}\right)+\cdots. 
\end{align}}
As can be seen in \Fig{asymplambda}, this approximation does well for small $\pran$ and $\tau$ for $r<(\pran,\tau)$.  
Note that although the only illustrated cases are those with $\pran=\tau$, we have tested cases down to $\phi\sim10^{-2}$.  

We may then use Equations \Eqff{nuapprox} and \Eqff{nuapproxchem} to estimate the thermal and compositional Nusselt numbers.  
Keeping only the leading order terms yields
\add{\begin{align}
\lEq{asympnusselt1}
\nut-1&\approx{C}^2\left(\pran\left(1+\phi\right)-\pran^{3/2}\frac{1+\phi^2}{\sqrt{1+\phi}}+\cdots\right), \\
\num-1&\approx{C^{2}}\left(\frac{\sqrt{1+\phi}}{\phi\sqrt{\pran}}-\frac{\phi}{1+\phi}+\cdots\right).
\end{align}}
This too is compared to the numerical solution in \Fig{asympnusselt}.

\subsection{Second regime: $(\pran,\tau)\ll{r}\ll1$}

In this regime, since $(\pran,\tau) \ll r $, we expand \Equation{cubic} and \Equation{quadratic} first in $\pran$ and then in $r$. Inspection of the numerical solutions suggests that $\lambda$ is proportional to $\pran$, while $l^2$ is more-or-less independent of it. Seeking solutions of \Equations{cubic} and \Eqff{quadratic} of the form $\lambda = \pran \hat \lambda$, to the lowest order in $\pran$, yields
\add{\begin{align}
	l^2 \hat{\lambda}^2+\left[l^4\left(1+\phi\right)+1\right]\hat{\lambda}+l^6\phi+l^2\phi\frac{r-1}{r}&=0\mbox{, and} \\
	\lEq{secondquadratic}\hat{\lambda}^2+2l^2\left(1+\phi\right)\hat{\lambda}+3l^4\phi+\phi\frac{r-1}{r}&=0.	
\end{align}}
\add{Eliminating the $\hat{\lambda}^2$ term yields}
\begin{equation}
\hat \lambda \left[ l^4(1+\phi)-1\right]+2l^6\phi=0 \mbox{  ,}
\end{equation}
which can be easily solved to find 
\begin{equation} 
\hat \lambda = -\frac{2l^6\phi}{l^4(1+\phi)-1} \mbox{  .}
\label{eq:hatlambda} 
\end{equation}
Plugging this into \Equation{secondquadratic}\del{, keeping only the lowest order terms in $\pran$,} and collecting the results in terms of $l$ yields 
\begin{equation} 
- l^{12} (1-\phi)^2 + l^8 (1+\phi) \left[ \phi - 1 - \frac{1+\phi}{r} \right]  + l^4 \left[ 1 - 2\phi + \frac{2}{r} (1+\phi) \right] + \frac{r-1}{r} = 0 
\lEq{cubicl4}
\end{equation}
which is a cubic equation for $l^4$. Unfortunately, this cubic has no simple solution, so we look for an expansion in the other asymptotic variable, $r$. Inspection of the numerical solutions suggests that the expansion needs to be in terms of $r^{1/2}$, so we set $l^4=l_0^4+l_1^4\sqrt{r}+l_2^4 r + \cdots$. Using this ansatz in \Equation{cubicl4} yields, to lowest order, 
\begin{equation}
l_0^8(1+\phi)^2 -2  l_0^4  (1+\phi) + 1 = 0,
\end{equation}
which we can solve to find $l_0^4=1/(1+\phi)$. This, on its own, causes $\hat \lambda$ to diverge (see \Equation{hatlambda}), so we need to go to the next order in $\sqrt{r}$. \add{By choosing the growing solution,} we then find that 
\begin{equation}
l^4=\frac{1}{1+\phi} - 2\frac{\sqrt{r\phi}}{(1+\phi)^{5/2}}+\cdots ,
\end{equation}
which eventually yields 
\add{\begin{equation}
\lEq{asymp2}
\lambda =\pran\sqrt{\frac{\phi}{r}}-\pran\sqrt{1+\phi}+\cdots ,
\end{equation}}
keeping the first two terms in the expansion only. As expected from the numerical solutions discussed above, $\lambda$ scales like $r^{-1/2}\pran $ to lowest order. 

As in the previous section, we use \Equation{nuapprox} and \Equation{nuapproxchem} along with these expansions to find approximate expressions for $\nut$ and $\num$:
\add{\begin{align}
\lEq{asympnusselt2}
\nut-1&\approx{C^2}\pran^2\left(\phi\frac{1+\phi}{r}-2\sqrt{\frac{\phi+\phi^2+\phi^3}{r\left(1+\phi\right)}}+\cdots\right)\mbox{, and} \\
\num-1&\approx{C^2}\left(\sqrt{\frac{1+\phi}{r\phi}}-\frac{1+2\phi+2\phi^2}{\phi\left(1+\phi\right)}+\cdots\right).
\end{align}}

We compare the results of the asymptotic expansion to the numerical solution in \Fig{asymplambda} and \Fig{asympnusselt}, where we look at $\lambda$ and $\num-1$, respectively.  
We choose to set the transition between the two regimes at $r=(\tau,\pran)$.  
For those simulations where $\tau$ is several orders of magnitude smaller than $\pran$, we recommend choosing this point at $r=\tau$.  
It is clear that the asymptotic expansion does an excellent job of reproducing the true numerical solution in the limit considered.  
We also see that, as suggested by the numerical data, $\num$ only depends on $\phi$ rather than on $\pran$ or $\tau$ individually as was noticed by \citet{Traxler2011a}.  

\begin{figure}
	\epsscale{0.5}
\includegraphics[width=\wid,angle=\rot]{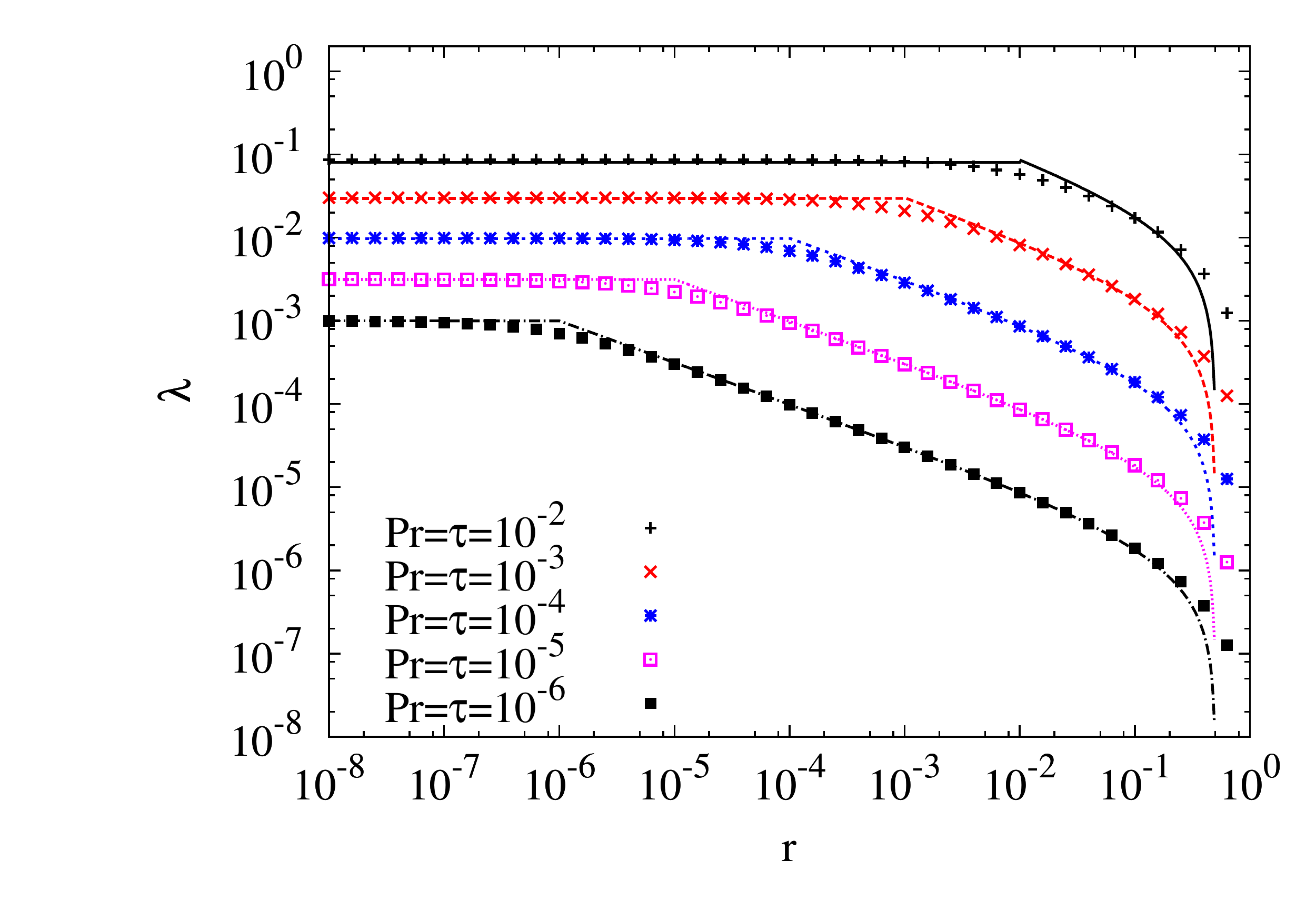}
\caption{\lFig{asymplambda} Comparison of the asymptotic expansions in Equations \Eqff{asymp1} and \Eqff{asymp2} to the numerical solution of $\lambda$ for small $r$ \del{using $C=7$} keeping up to the second order. The asymptotic expansion switches between two regimes of low $r$, $r\gg(\tau,\pran)$ and $r\ll(\tau,\pran)$, at $r=(\tau,\pran)$. The data points represent the numerical solution and the curves represent the asymptotic expansion. Note that the aysmptotic expansion fits the data remarkably well in the respective limits considered.}
\end{figure}

\begin{figure}
	\epsscale{0.5}
\includegraphics[width=\wid,angle=\rot]{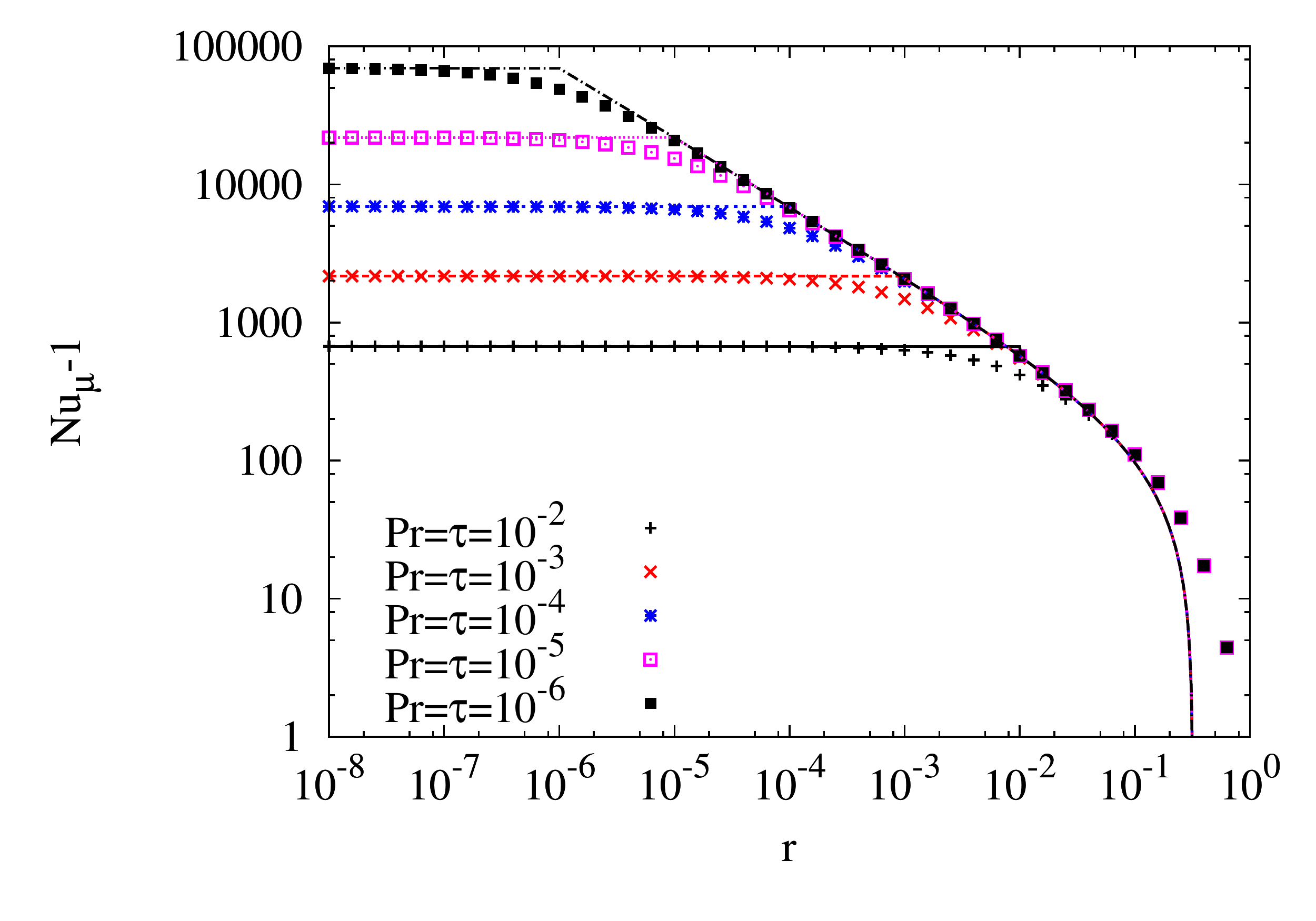}
\caption{\lFig{asympnusselt} Comparison of the asymptotic expansions in Equations \Eqff{asympnusselt1} and \Eqff{asympnusselt2} to the numerical solution of $\num-1$ from \Equation{nuapproxchem} for small $r$ using $C=7$ keeping up to the second order. Note that this retains the same quality of fit as \Fig{asymplambda}.}
\end{figure}

\subsection{Third regime: $r  \rightarrow 1$}

Equations \Eqff{hatlambda} and \Eqff{cubicl4} were derived without 
any assumptions on the value of $r$, aside from the fact that it needs to be much larger than $\pran$ and $\tau$. 
We can therefore use them directly to study the limit of $r  \rightarrow 1$. To do so, we set $\epsilon = 1-r $, and rewrite  \Equation{cubicl4} in terms of the small parameter $\epsilon$: 
\add{\begin{equation}
l^{12}  (1-\phi)^2\left(\epsilon-1\right) - l^8(1+\phi)  \left[  2 + \epsilon(\phi-1))   \right]  + l^4 \left[3+\epsilon\left(2\phi-1\right)\right] - \epsilon = 0.
\label{eq:cubicl4r1}
\end{equation}}
Inspection of the solutions for $l^4$ in the limit of small $\epsilon$ suggests the expansion  $l^4 = \epsilon  ( l_0^4 +\epsilon l_1^4 + \cdots ) $. Plugging this into \Equation{cubicl4r1} then yields, to lowest order,  
\begin{equation}
\lEq{largel}
l^4 = \frac{1}{3} (1-r)+\cdots.
\end{equation}
Plugging this expression into \Equation{hatlambda}, we get 
\add{\begin{equation}
\lEq{largelambda}
\lambda = 2 \pran \phi \left[ \frac{1}{3} (1-r) +\cdots\right]^{3/2}\left[ 1-  \frac{1+\phi}{3} (1-r) +\cdots\right]^{-1}.
\end{equation}}

As before, we use Equations \Eqff{largel} and \Eqff{largelambda} with Equations \Eqff{nuapprox} and \Eqff{nuapproxchem} to find expansions in $\pran$ and $r$ for $\nut$ and $\num$ in the limits $\pran\to0$ and $r\to1$.  
We find
\add{\begin{align}
\lEq{largenu}
\nut-1&\approx{C^{2}}\pran^{2}\left(\frac{4}{9}\phi^{2}\left(1-r\right)^{2}-\frac{8}{81}\phi^{2}\left(\phi-8\right)\left(1-r\right)^{3}\cdots\right), \nonumber \\
\num-1&\approx{C^{2}}\left(\frac{4}{9}\left(1-r\right)^{2}-\frac{8}{81}\left(\phi-5\right)\left(1-r\right)^{3}+\cdots\right).
\end{align}}

We compare the asymptotic expressions to the numerical solutions for $\lambda$ and $\num$ in \Fig{largeasymplambda} and in \Fig{largeasympnusselt}.  
The expansion clearly works best for $r\gtrsim0.5$.  
Note that these asymptotic expansions are designed to work only when $\phi\sim1$ and should be used with caution if $\phi\ll{r}$.  

\begin{figure}
	\epsscale{0.5}
\includegraphics[width=\wid,angle=\rot]{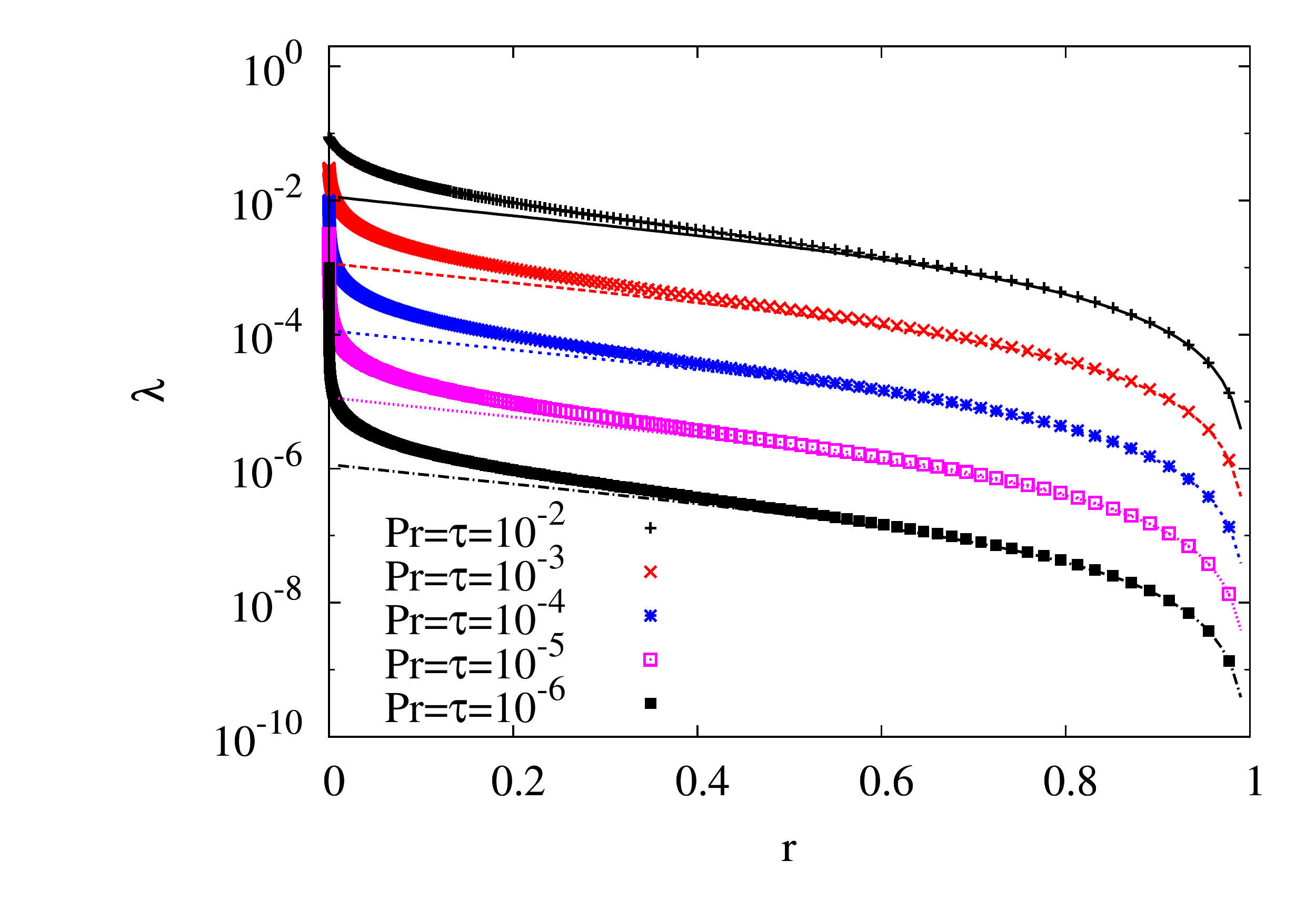}
\caption{\lFig{largeasymplambda} Comparison of the asymptotic expansion in \Equation{largelambda} to the numerical solution of $\lambda$ as $r\to1$. The data points represent the numerical solution and the curves represent the asymptotic expansion. Note that the aysmptotic expansion fits the data well for its intended regime.}
\end{figure}

\begin{figure}
	\epsscale{0.5}
\includegraphics[width=\wid,angle=\rot]{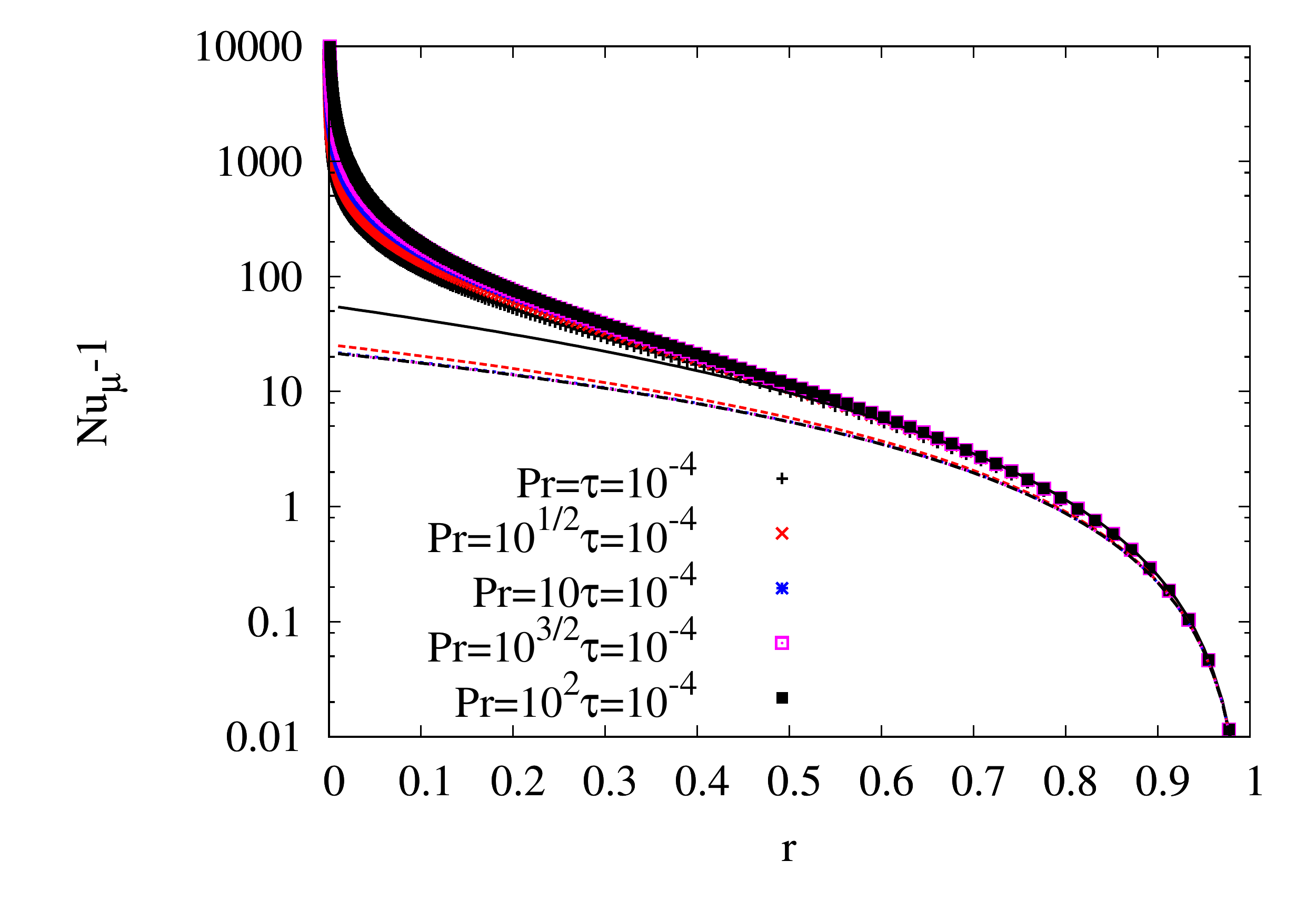}
\caption{\lFig{largeasympnusselt} Comparison of the asymptotic expansion in \Equation{largenu} to the numerical solution of $\num-1$ from \Equation{nuapproxchem} as $r\to1$ with $C=7$. Note that this retains the same quality of fit as \Fig{largeasymplambda} \add{for $\phi=1$, but still remains within a factor of 2 down to $r=0.5$ for all tested values of $\phi$}.}
\end{figure}


\begin{thebibliography}{36}
\expandafter\ifx\csname natexlab\endcsname\relax\def\natexlab#1{#1}\fi

\bibitem[{Baines \& Gill(1969)}]{Baines1969}
Baines, P.~G. \& Gill, A.~E. 1969, Journal of Fluid Mechanics, 37, 289

\bibitem[{Calzavarini {et~al.}(2005)Calzavarini, Lohse, Toschi, \&
  Tripiccione}]{Calzavarini2005}
Calzavarini, E., Lohse, D., Toschi, F., \& Tripiccione, R. 2005, Physics of
  Fluids, 17, 055107

\bibitem[{Chapman \& Cowling(1970)}]{Chapman1970}
Chapman, S. \& Cowling, T. 1970, {The mathematical theory of non-uniform gases.
  an account of the kinetic theory of viscosity, thermal conduction and
  diffusion in gases} (Cambridge University Press)

\bibitem[{Charbonnel(1994)}]{Charbonnel1994}
Charbonnel, C. 1994, Astronomy and Astrophysics (ISSN 0004-6361), vol. 282, no.
  3, p. 811-820

\bibitem[{Charbonnel \& Zahn(2007)}]{Charbonnel2007}
Charbonnel, C. \& Zahn, J.-P. 2007, \aap, 467, L15

\bibitem[{Denissenkov(2010)}]{Denissenkov2010}
Denissenkov, P. 2010, \apj, 723, 563

\bibitem[{Eggleton {et~al.}(2006)Eggleton, Dearborn, \&
  Lattanzio}]{Eggleton2006}
Eggleton, P.~P., Dearborn, D. S.~P., \& Lattanzio, J.~C. 2006, Science (New
  York, N.Y.), 314, 1580

\bibitem[{Fischer \& Valenti(2005)}]{Fischer2005}
Fischer, D.~A. \& Valenti, J. 2005, The Astrophysical Journal, 622, 1102

\bibitem[{Garaud(2011)}]{Garaud2011}
Garaud, P. 2011, The Astrophysical Journal, 728, L30

\bibitem[{Garaud {et~al.}(2010)Garaud, Ogilvie, Miller, \&
  Stellmach}]{Garaud2010}
Garaud, P., Ogilvie, G.~I., Miller, N., \& Stellmach, S. 2010, Monthly Notices
  of the Royal Astronomical Society, 407, 2451

\bibitem[{Gilroy(1989)}]{Gilroy1989}
Gilroy, K.~K. 1989, The Astrophysical Journal, 347, 835

\bibitem[{Kato(1966)}]{Kato1966}
Kato, S. 1966, \pasj, 18, 374

\bibitem[{Kippenhahn {et~al.}(1980)Kippenhahn, Ruschenplatt, \&
  Thomas}]{Kippenhahn1980}
Kippenhahn, R., Ruschenplatt, G., \& Thomas, H.-C. 1980, \aap, 91, 175

\bibitem[{Kippenhahn \& Weigert(1990)}]{Kippenhahn1990}
Kippenhahn, R. \& Weigert, A. 1990, {Stellar structure and evolution},
  Astronomy and astrophysics library (Springer)

\bibitem[{Kunze(2003)}]{Kunze2003}
Kunze, E. 2003, Progress In Oceanography, 56, 399

\bibitem[{Mirouh {et~al.}(2012)Mirouh, Garaud, Stellmach, Traxler, \&
  Wood}]{Mirouh2012}
Mirouh, G.~M., Garaud, P., Stellmach, S., Traxler, A.~L., \& Wood, T.~S. 2012,
  The Astrophysical Journal, 750, 61

\bibitem[{Radko(2003)}]{Radko2003}
Radko, T. 2003, Journal of Fluid Mechanics, 497, 365

\bibitem[{Radko \& Smith(2012)}]{Radko2012}
Radko, T. \& Smith, D. 2012, Journal of Fluid Mechanics, 692, 5

\bibitem[{Rosenblum {et~al.}(2011)Rosenblum, Garaud, Traxler, \&
  Stellmach}]{Rosenblum2011}
Rosenblum, E., Garaud, P., Traxler, A., \& Stellmach, S. 2011, \apj, 731, 66

\bibitem[{Santos {et~al.}(2001)Santos, Israelian, \& Mayor}]{Santos2001}
Santos, N.~C., Israelian, G., \& Mayor, M. 2001, Astronomy and Astrophysics,
  373, 1019

\bibitem[{Schmitt(1979)}]{Schmitt1979b}
Schmitt. 1979, J. Mar. Res., 37, 419

\bibitem[{Schmitt(1983)}]{Schmitt1983}
Schmitt, R.~W. 1983, Physics of Fluids, 26, 2373

\bibitem[{Schmitt {et~al.}(2005)Schmitt, Ledwell, Montgomery, Polzin, \&
  Toole}]{Schmitt2005}
Schmitt, R.~W., Ledwell, J.~R., Montgomery, E.~T., Polzin, K.~L., \& Toole,
  J.~M. 2005, Science, 308

\bibitem[{Spiegel \& Veronis(1960)}]{Spiegel1960}
Spiegel, E. \& Veronis, G. 1960, \apj, 131, 442

\bibitem[{Stellmach {et~al.}(2011)Stellmach, Traxler, Garaud, Brummell, \&
  Radko}]{Stellmach2011}
Stellmach, S., Traxler, A., Garaud, P., Brummell, N., \& Radko, T. 2011,
  Journal of Fluid Mechanics, 677, 554

\bibitem[{Stern(1969)}]{Stern1969b}
Stern, M. 1969, Journal of Fluid Mechanics, 35, 209

\bibitem[{Stern(1960)}]{Stern1960}
Stern, M.~E. 1960, Tellus, 12, 172

\bibitem[{Stern {et~al.}(2001)Stern, Radko, \& Simeonov}]{Stern2001}
Stern, M.~E., Radko, T., \& Simeonov, J. 2001, Journal of Marine Research, 59,
  355

\bibitem[{Stern \& Turner(1969)}]{Stern1969a}
Stern, M.~E. \& Turner, J.~S. 1969, Deep Sea Research and Oceanographic
  Abstracts, 16, 497

\bibitem[{Traxler {et~al.}(2011{\natexlab{a}})Traxler, Garaud, \&
  Stellmach}]{Traxler2011a}
Traxler, A., Garaud, P., \& Stellmach, S. 2011{\natexlab{a}}, \apjl, 728, L29

\bibitem[{Traxler {et~al.}(2011{\natexlab{b}})Traxler, Stellmach, Garaud,
  Radko, \& Brummell}]{Traxler2011b}
Traxler, A., Stellmach, S., Garaud, P., Radko, T., \& Brummell, N.
  2011{\natexlab{b}}, Journal of Fluid Mechanics, 677, 530

\bibitem[{Turner(1967)}]{Turner1967}
Turner, J. 1967, Deep Sea Research and Oceanographic Abstracts, 14, 599

\bibitem[{Ulrich(1972)}]{Ulrich1972}
Ulrich, R. 1972, \apj, 172, 165

\bibitem[{Vauclair(2004)}]{Vauclair2004}
Vauclair, S. 2004, The Astrophysical Journal, 605, 874

\bibitem[{Vauclair \& Th\'{e}ado(2012)}]{Vauclair2012}
Vauclair, S. \& Th\'{e}ado, S. 2012, \apj, 753, 49

\bibitem[{Wood {et~al.}(2013)Wood, Garaud, \& Stellmach}]{Wood2013}
Wood, T.~S., Garaud, P., \& Stellmach, S. 2013, 19

\end{thebibliography}
\end{document}